\documentclass{nature}
\usepackage[numbers]{natbib}
\usepackage{xcolor}
\usepackage{hyperref}
\definecolor{docnotelinkcolor}{RGB}{0,0,255}
\usepackage[T1]{fontenc}
\usepackage[left]{lineno}
\usepackage{amsthm,amsmath,amssymb}
\usepackage{mathrsfs}
\usepackage{overpic}
\usepackage[T1]{fontenc}
\usepackage[numbers]{natbib}
\usepackage{ulem}
\usepackage[caption = false]{subfig}
\usepackage{academicons}
\usepackage{amsmath, amssymb}
\usepackage{graphicx}
\usepackage{hyperref}
\usepackage{threeparttable}
\usepackage{xcolor}
\usepackage{xcolor}

\newenvironment{redsubsection}[1]{%
  \subsection{#1}%
  \begingroup
  \color{black}%
  \everymath{\color{black}}%
  \everydisplay{\color{black}}%
}{%
  \par
  \endgroup
}
\usepackage{ulem}
\usepackage{rotating}
\usepackage{multibib}

\newcommand{\vsim}{\mathrel{\rotatebox{90}{$\sim$}}}
\usepackage{caption}
\def\emph#1 {\textit{ #1 } }
\newcommand{\apj}{Astrophys. J.}

\newcommand{\apjs}{Astrophys. J. Supp.}
\newcommand{\araa}{Annu. Rev. Astron. Astrophys.}
\newcommand{\mnras}{Mon. Not. R. Astron. Soc.}
\newcommand{\apjl}{Astrophys. J. Let.}
\newcommand{\aap}{Astron. Astrophys.}

\newcommand{\nat}{Nature}
\newcommand{\physrep}{Physics Reports}

\title{Relativistic \(^{56}\text{Ni}\) Decay Lines in GRB 221009A}

\author{
Rahim Moradi$^{1}$\thanks{E-mail: rmoradi@ihep.ac.cn},
Emre S. Yorgancioglu$^{1,2}$\thanks{E-mail: emre@ihep.ac.cn},
Shao-Lin Xiong$^{1}$\thanks{E-mail: xiongsl@ihep.ac.cn},
Yan-Qiu Zhang$^{1,2,3}$
Shuang-Nan Zhang$^{1,2}$,
Roland Diehl$^{4}$,
Yu Wang$^{5,6,7,8}$\thanks{E-mail: yu.wang@icranet.org}
}

\begin{document}

\maketitle

\begin{affiliations}

 \item State Key Laboratory of Particle Astrophysics, Institute of High Energy Physics, Chinese Academy of Sciences, Beijing 100049, China
 \item University of Chinese Academy of Sciences, Chinese Academy of Sciences, Beijing 100049, China
 \item School of Physics and Electronic Science, Guizhou Normal University, Guiyang 550001, China
 \item Max Planck Institut f\"ur extraterrestrische Physik 
and Technical University (TUM), 85748, Garching, Germany
 \item ICRA, Dipartamento di Fisica, Sapienza Universit\`a  di Roma, Piazzale Aldo Moro 5, I-00185 Rome, Italy
 \item ICRANet, Piazza della Repubblica 10, I-65122 Pescara, Italy
  \item ICRANet-AI, Brickell Avenue 701, Miami, FL 33131, USA
 \item INAF - Osservatorio Astronomico d'Abruzzo, Via M. Maggini snc, I-64100, Teramo, Italy

 \end{affiliations}

\begin{abstract}

{Long Gamma Ray Bursts are thought to originate from the core collapse of massive stars that give rise to energetic broad-lined Type Ic supernovae. The brightest burst ever recorded, GRB 221009A, has been linked to a broad-lined Type Ic supernova through late-time observations by the James Webb Space Telescope. An emission line evolving from $\sim$37 to $\sim$6~MeV is detected during the prompt phase. We propose that this time-evolving line is consistent with Doppler-boosted radioactive decay of nickel synthesized in the associated supernova and entrained in the relativistic jet, corresponding to the boosted 158~keV decay branch. We also report evidence for an additional higher-energy excess near $\sim$24~MeV at 290--300~s, detected at moderate statistical significance and consistent with the boosted 270~keV decay branch. The observed kinematics and flux evolution are compatible with expectations from radioactive decay, providing direct spectroscopic evidence linking prompt emission to supernova nucleosynthesis.
}
\end{abstract}

\section*{Introduction}

Long-duration gamma-ray bursts (LGRBs) mark the deaths of massive stars, and in a subset of cases are accompanied by broad-lined Type Ic supernovae (SNe), as a relativistic jet pierces the progenitor envelope \citep{2003Natur.423..847H,galama1998unusual,2006RPPh...69.2259M,2018pgrb.book.....Z}. These GRB-SNe lack hydrogen and helium lines and exhibit broad absorption features from high-velocity ejecta (v $\lesssim$ 0.1 c) powered by the nascent black hole or neutron star \citep{2003fthp.conf...87W,2001ApJ...550..410M,2017AdAst2017E...5C}. They synthesize 0.1–0.5 M$_\odot$ of ${}^{56}\mathrm{Ni}$, whose decay to ${}^{56}\mathrm{Co}$ produces an optical ``bump'' at $\sim$ 13 days with an isotropic luminosity of L $\sim 10^{43}\,\mathrm{erg\,s}^{-1}$ \citep{2017AdAst2017E...5C,2023ApJ...955...93A}. Although first solidified by GRB 980425/SN 1998bw \citep{galama1998unusual}, spectroscopic confirmation remains scarce: fewer than 1\% of GRBs show SN signatures, {\color{black} mainly due to redshift (z) and observational limitations; in rare cases where nearby (z $\lesssim$ 1) long GRBs have been followed with adequate optical sensitivity, an associated SN has nearly always been detected} \citep{2017AdAst2017E...5C}. Early SN detection, as achieved for the low-luminosity GRB 171205A/SN 2017iuk—with a hot cocoon outshining the SN until $\lesssim$ 3 days—offers critical insight into jet–ejecta interactions and nucleosynthesis \citep{2019Natur.565..324I}. Such observations are essential to constrain the physics of GRB central engines and the role of radioactive material in shaping their multi-wavelength light curves.

On October 9, 2022, GRB 221009A was detected by multiple gamma-ray instruments{, including the Fermi Gamma-ray Burst Monitor (GBM), the Swift Burst Alert Telescope (BAT), the Gravitational-wave High-energy Electromagnetic Counterpart All-sky Monitor (GECAM-C), the Konus--Wind instrument, and the Hard X-ray Modulation Telescope (Insight-HXMT)}, with a reported fluence of 0.2\,$\mathrm{erg\,cm^{-2}}$ \citep[e.g.][]{2023arXiv230301203A, frederiks2023properties}. With the host galaxy's redshift of $z = 0.151$, GECAM-C accurate measurements of GRB 221009A yield an {exceptionally high} intrinsic isotropic energy of $E_{\mathrm{iso}} \sim 1.5 \times 10^{55}$ erg and a peak luminosity of $L_{\mathrm{p, iso}} \sim 10^{54}\, \mathrm{erg\,s^{-1}}$ {\color{black} in the
hard X-ray to soft gamma-ray band from 10 keV to $\sim$ 6 MeV}, making it the most luminous GRB observed to date \citep{2023arXiv230301203A}. Recent observations with the James Webb Space Telescope (JWST) obtained 168–170 days (in the rest frame) after the GRB trigger revealed spectra consistent with a  1998bw-like SN, showing key features such as the Ca II near-infrared triplet and O I lines, typical of core-collapse supernovae. The inferred mass of ${}^{56}\mathrm{Ni}$ in the ejecta was 0.09 $M_\odot$, consistent with typical GRB-associated SNe \citep{blanchard2024jwst}. 

In addition to its {extreme} brightness \citep{2023arXiv230301203A} and the first {tera–electron volt (TeV)} afterglow measured by LHAASO \citep{2023SciA....9J2778C}, two independent studies reported narrow MeV emission lines in the prompt spectrum of GRB 221009A \citep{2024SCPMA..6789511Z,2024Sci...385..452R}. Fermi/GBM data, unusable during peak saturation, revealed significant emission line features near 10 MeV in the 280-320 s post-trigger window \citep{2024Sci...385..452R}. However, a joint analysis of GECAM-C and Fermi/GBM identified {a time-evolving sequence of} emission lines whose centroids drift from 37 MeV to 6 MeV as a time evolution with power law index of $\approx-1$, maintain a constant broadening of $\sim$ 10\%, and exhibit a flux decay by a power law index $k \sim -2$ \citep{2024SCPMA..6789511Z}.

{A confirmed detection of an emission line after decades of GRB observations} raises ongoing debate about its origin \cite[e.g.][]{2024ApJ...968L...5W, 2025ApJ...983L..33Z}. This line is often attributed to pair annihilation within the relativistic jet \cite{2024SCPMA..6789511Z, 2024Sci...385..452R, 2024ApJ...973L..17Z, yi2024robust, pe2024physical,liu2025production}, the excitation of hydrogen-like high-Z ions \cite{2024ApJ...968L...5W}, or the 2.223 MeV gamma-rays following neutron capture with protons  \citep{2025ApJ...983L..33Z}. In the pair annihilation scenario, the observed temporal decay with a power-law index of $k \sim -2$ contradicts the $k = -3$ predicted by high-latitude emission (HLE) \citep{2024SCPMA..6789511Z}, suggesting either parameter fine-tuning \cite{2024ApJ...973L..17Z} or a rejection of the HLE geometry. Furthermore, if the broadband HLE emission dominates the prompt spectrum, it would obscure the spectral line, as the temporal evolution of the broadband emission should match that of the line, which is not observed \cite{2024Sci...385..452R}.

We here propose that the observed MeV emission lines arise from the Doppler-boosted decay of $^{56}\mathrm{Ni}$ synthesized during the explosion and entrained in the relativistic jet, \textcolor{black}{consistent with the 158~keV decay branch in the laboratory
frame. A higher-energy excess is further found to be consistent with the
270~keV decay branch.} {In this work, we model the time-dependent spectral and luminosity evolution of the line and derive constraints on the jet dynamics and entrained nickel mass.} This result provides direct observational evidence linking prompt
GRB emission with supernova nucleosynthesis during the early phases of the burst.

\section*{Results}

\subsection{Emission Lines from Relativistic \texorpdfstring{\({}^{56}\mathrm{Ni}\)}{56Ni} Decay}

\begin{figure}[h]
    \centering
    \includegraphics[width=0.7\textwidth]{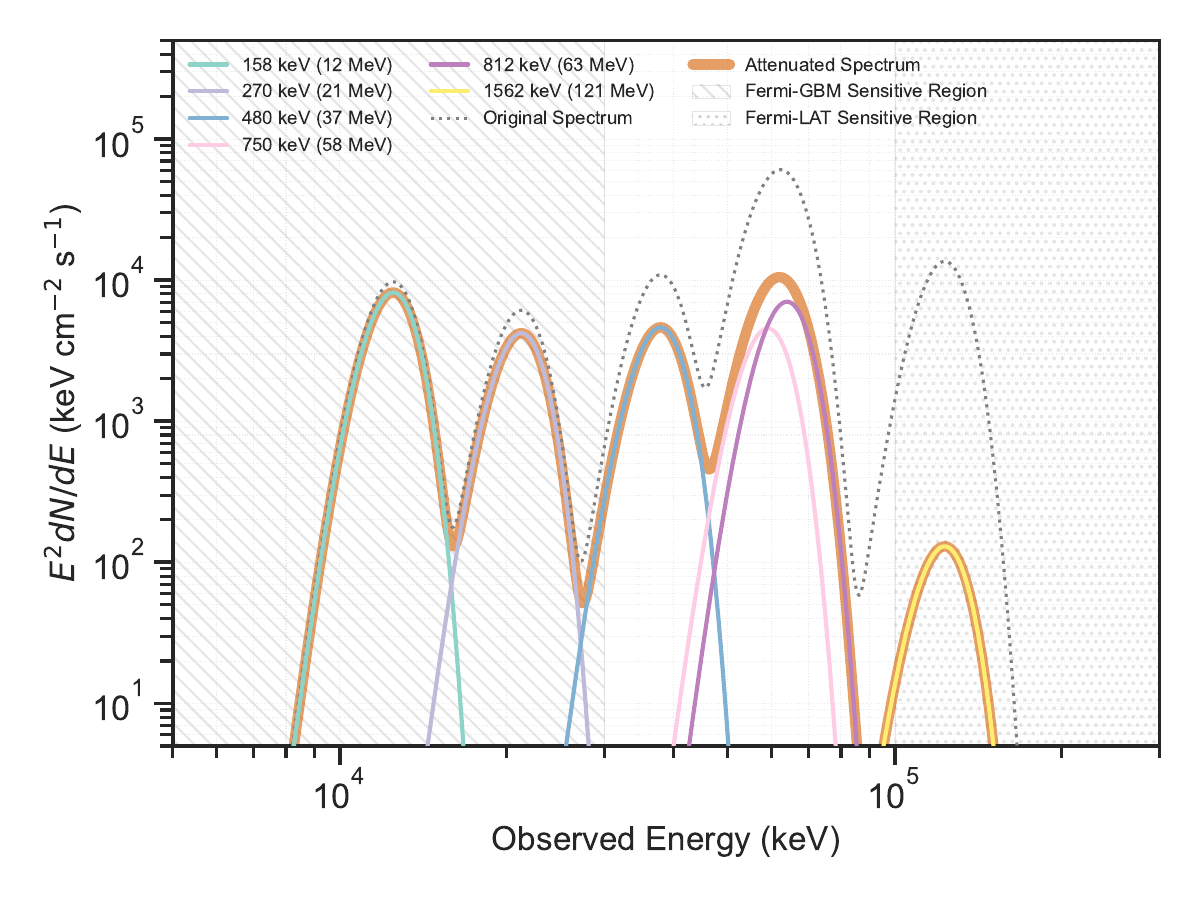}
    \caption{\textcolor{black}{{\textbf{Predicted observable spectrum of Nickel lines}}. This includes $\gamma\gamma$ annihilation assuming the internal shock occurs at $R=10^{14}$ cm at time 290-300 s. The plot shows the $E^2 \mathrm{d}N/\mathrm{d}E$ spectrum after applying the  attenuation to each individual $^{56}\mathrm{Ni}$ line component due to $\gamma\gamma$ collisions. The total attenuated spectrum (thick solid orange line) is compared to the intrinsic spectrum (dotted grey line). Individual attenuated line components are also shown (colored solid lines). Spectra are normalized such that the attenuated 158 keV (observed 12 MeV) line peak aligns with the observed peak. Hatched and dotted regions indicate the sensitive ranges of Fermi-GBM and Fermi-LAT, respectively. Note the significant suppression of the 1562 keV line (attenuation factor $\sim 0.007$). If the internal shock occurs at a smaller radius, the attenuation, especially for higher energy lines, will be stronger.}}
    \label{fig:nickel-decay-lines}
\end{figure}

The decay scheme of ${}^{56}\mathrm{Ni}\rightarrow{}^{56}\mathrm{Co}$ in the laboratory is dominated by a transition at $E_{\gamma}=158.38\ \mathrm{keV}$ ($3^{+}\rightarrow4^{+}$) that carries a branching intensity of $I_{\gamma} = 98.8\%$; this intensity represents the fraction of total decays in which the respective gamma-ray photon is emitted. Thus almost every ${}^{56}\mathrm{Ni}$ nucleus that decays emits a 158 keV photon. Additional branches with significant intensities are 811.85 keV ($2^{+}\rightarrow3^{+}$, $I_{\gamma}=86.0\%$), 749.95 keV ($1^{+}\rightarrow2^{+}$, $I_{\gamma}=49.5\%$), 480.44 keV ($0^{+}\rightarrow2^{+}$, $I_{\gamma}=36.5\%$), 269.50 keV ($1^{+}\rightarrow0^{+}$, $I_{\gamma}=36.5\%$), and 1561.80 keV ($1^{+}\rightarrow3^{+}$, $I_{\gamma}=14.0\%$) \cite{sur1990reinvestigation, 1994ApJS...92..527N}; see Table.~\ref{tab:ni_decay}. Previous work {\color{black} on GRB 221009A} \cite{2024SCPMA..6789511Z} has shown an emission line in the spectrum evolving from $\sim 37$ MeV at 250 s to 6 MeV at 350 s, which in this article is interpreted as the Doppler boosted 158 keV line from the decay of  ${}^{56}\mathrm{Ni}$ entrained in the relativistic jet possibly within the {collapsar} model for long-duration GRBs and their associated broad-lined Type Ic SNe. In this framework, a rapidly rotating massive star undergoes core collapse to form a black hole surrounded by a hot, dense accretion disk. Accretion energy, via neutrino annihilation or magnetic processes, launches relativistic jets through the stellar envelope \cite{1993ApJ...405..273W,1999ApJ...524..262M,Woosley:2002cp,2006ARA&A..44..507W}. Temperatures in the disk (\(T>10^9\) K) enable explosive nucleosynthesis, generating heavy elements like \({}^{56}\mathrm{Ni}\), which are then mixed into jets and a surrounding cocoon through hydrodynamic instabilities \cite{2003A&A...408..621K,2010ApJ...714.1371H,2006ApJ...647.1255N,MacFadyen:1998vz}. 

The second and third most intense lines, 811.85 keV and 749.95 keV,  are a factor of five more energetic than the 158 keV line. For observer times $t<260\ \mathrm{s}$, the 158 keV photons are Doppler shifted above 20 MeV. In the same time interval, the $\sim 800$ keV photons would appear above 100 MeV, within the LAT bandwidth. However, due to instrumental constraints, Fermi-LAT only began recording data at time after 290 s, when the 158 keV line redshifted to $\simeq10\ \mathrm{MeV}$ and the 800 keV lines  to $\simeq50\ \mathrm{MeV}$, near the edges of the response curves of both Fermi-GBM and Fermi-LAT. The reduced effective area at these energies limits the sensitivity to the 800 keV features. 

Both the 480 keV and 270 keV decay branches have intensities of about 40\%. Initially (t $\lesssim$ 290 s), they Doppler-shift to energies above the GBM’s sensitive range.  By t $\approx$ 290 s, the 270 keV line falls squarely into the 20–40 MeV window, ideal for GBM detection, whereas the 480 keV line still sits at the very upper edge of the instrument’s response, where sensitivity is poor.  At later times, the intrinsic line flux fades and the line energies drift toward the peak of the prompt continuum, further reducing their contrast against the background and making them harder to detect. 

\textcolor{black}{Figure~\ref{fig:nickel-decay-lines} illustrates the expected $\gamma\gamma$-annihilation attenuation together with the instrumental sensitivity ranges of Fermi-GBM and Fermi-LAT. It shows that the Doppler-boosted 158 and 270~keV lines are the most likely to be detected, while higher-energy branches suffer stronger attenuation and fall partially outside the effective response windows. Detailed calculations and discussions are provided in the Methods.}

\begin{figure}[h]
    \centering
    \includegraphics[width=0.7\textwidth]{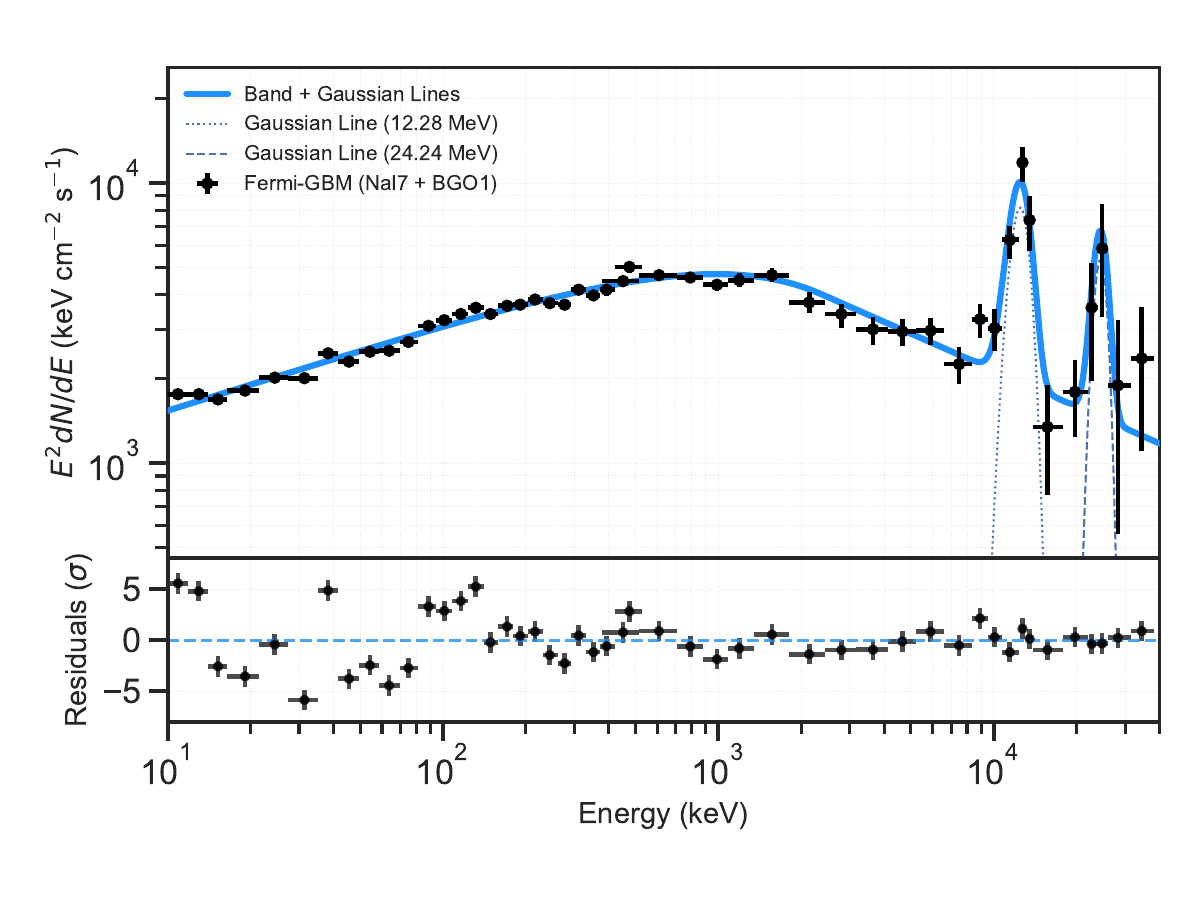}
    \caption{{\textbf{The $E^2 dN/dE$ spectrum of GRB 221009A observed by Fermi-GBM (NaI7+BGO1)}}. The spectrum is integrated from 290 s to 300 s and the data are represented by the black points. Error bars represent 1 $\sigma$ errors. {\color{black} This time interval (290--300~s) is when the second line significance reaches maximal, resulting in smaller statistical errors than in the broader, data-driven time intervals selected by Zhang et al \cite{2024SCPMA..6789511Z}, and Ravasio et al. (2024) \citep{2024Sci...385..452R}.} The solid blue line shows a model comprising a Band function plus two Gaussian \textcolor{black}{components}  centered at 12.28 MeV (dotted) and at 24.24 MeV (dashed). The lower panel displays the residuals (data - model) in units of standard deviation ($\sigma$).}
    \label{fig:270line}
\end{figure}

The spectral fitting between 290 s and 300 s reveals two  Gaussian components centered at $\sim 12.22\pm0.03\ \mathrm{MeV}$ and $\sim 24.24\pm0.12\ \mathrm{MeV}$, see Figure \ref{fig:270line}. \textcolor{black}{At this time interval, the centroid of the previously reported time-evolving MeV line ($\sim$37$\rightarrow\sim$6 MeV) lies near $\sim$12 MeV; thus the $\sim$12 MeV component in Figure \ref{fig:270line} represents a snapshot of that same evolving line integrated over 290–300 s.} The 12 MeV line has been reported in \cite{2024Sci...385..452R, 2024SCPMA..6789511Z}, while \textcolor{black}{an additional spectral excess near $\sim$24~MeV is identified}  in this paper using a $C$-stat likelihood-ratio test, yielding $\Delta C = 8.93$, \textcolor{black}{and is therefore treated as a moderate-significance feature rather than a firm line detection}.
 Depending on the null hypothesis treatment, this corresponds to a Gaussian significance in the range $1.9\sigma$--$3.0\sigma$, providing moderate statistical support for the additional spectral component. Moreover, a Bayesian {Markov chain Monte Carlo (MCMC)} model comparison implies a Bayes factor of $\simeq 10$, that is, the two-line model is about ten times more likely than the one-line model (see Methods).

Assuming the 12 MeV feature matches the expected energy of the Doppler-boosted 158 keV transition, then the 24 MeV feature \textcolor{black}{is consistent with the} boosted 270 keV \textcolor{black}{decay branch}. {\color{black}The observed energy ratio of the two features, $E_{2}/E_{1} = 24.24/12.22 \approx 1.98$, exceeds the laboratory value of $270/158 \approx 1.71$ by approximately 16\%. \textcolor{black}{While a complete model of nuclear processes in a relativistic jet remains uncertain, ionization or intrinsic nuclear physics is unlikely to explain the mismatch because they do not shift the comoving rest energies obviously, a more direct explanation is energy-dependent radiative transfer, where Klein--Nishina suppression lowers the opacity at 270~keV so those photons decouple at a smaller radius than the 158~keV photons, inducing a modest Doppler factor difference and/or a small change in the effective last-scattering angle (see Methods).}

\textcolor{black}{
We emphasize that this behavior is not a defect of the model but a direct
consequence of its predictive nature: the same physical framework that explains
the observed time-evolving MeV emission line as the Doppler-boosted manifestation
of the $^{56}$Ni 158~keV decay branch generically predicts additional decay
branches whose observability depends on energy-dependent opacity, outflow
geometry, and instrumental response, thereby providing a testable and
falsifiable observational signature.
}

Internal conversion coefficients for the 158 keV and 270 keV transitions are $\sim10^{-2}$ and $\sim 10^{-3}$, respectively \citep{2008NIMPA.589..202K}, altering the photon ratio by only one percent for either neutral gas or highly ionized plasma. Both transitions are far below the pair-creation threshold for interactions with the abundant X-ray photons in the comoving frame, but they can interact with the scarcer high-energy tail, so $\gamma\gamma$ annihilation is weak. Our calculation (Section Methods) shows that at most tens of percent of the 158 keV photons and up to a factor of 1-2.5 times more of the 270 keV photons are annihilated. Additionally, the Klein–Nishina cross-section is still close to the Thomson value at these energies, Compton scattering attenuates the two transitions almost equally. Therefore, a modest reduction of the 270 keV to 158 keV photon ratio relative to the laboratory value is  expected, in agreement with the measurement.

The 1561.80 keV branch has an energy one order of magnitude above that of the 158 keV line. At $t \sim 290\ \mathrm{s}$, it is Doppler shifted to $\sim 100\ \mathrm{MeV}$, which lies within the nominal Fermi-LAT range. In contrast to the lower-energy lines, the boosted 1562 keV photons exceed the pair-creation threshold against the dense X-ray field from the prompt emission, and almost all are quenched by $\gamma\gamma$ annihilation. Consequently no corresponding feature will be present in the Fermi-LAT data, consistent with the observation, see Methods for details. {\color{black} We also assess whether a $\gamma\gamma$ absorption trough can provide clues to the emission line origin. For the $E_{\rm em}=1561.8~\mathrm{keV}$ branch, the target photons that maximize the pair-production cross section have comoving energy $\sim 350~\mathrm{keV}$. In the observer frame this energy is Doppler-boosted to $\simeq 26~\mathrm{MeV}$ for $D\simeq 75$ during $t=290$–$300$s and to $\gtrsim 7~\mathrm{MeV}$ for the lowest $D\simeq 20$. At these MeV energies the GBM effective area is small and LAT has poor narrow-line sensitivity, which makes any $\gamma\gamma$ absorption imprint difficult to measure. For lower-energy decay lines, the corresponding target photon energy increases, placing it even further beyond the sensitive range of Fermi-GBM, and making such absorption features essentially undetectable.} 

\textcolor{black}{
We emphasize that the interpretation of the higher-energy feature is limited by
photon statistics and instrumental systematics in the brightest phases of
GRB~221009A. The interpretation of the $\sim24$~MeV excess is therefore tentative.
Importantly, the validity of the $^{56}$Ni-decay interpretation does not rely on
the detection of this second feature; the central result of this work is that
the observed time-evolving MeV emission line during the prompt phase, whose
centroid energy evolves from $\sim$37 to $\sim$6~MeV, is consistent with
Doppler-boosted radioactive decay of $^{56}$Ni (rest-frame 158~keV) synthesized
in the associated supernova and entrained in the relativistic jet. Confirmation
of additional decay branches must await future observations with improved photon
statistics and spectral resolution. More generally, the quantitative inferences
on jet dynamics and nickel mass rely on assumptions about the geometry and
radiative transfer at the last-scattering surface, which may be refined by future
observations and more detailed modeling.
}

\subsection{The Observed Luminosity from Nickel Decay}

We first calculate the luminosity in the jet’s rest frame. Radioactive \({}^{56}\mathrm{Ni}\) decays into \({}^{56}\mathrm{Co}\), releasing keV--MeV range gamma-ray photons \citep{1994ApJS...92..527N}, whose energies are listed in Table. \ref{tab:ni_decay}. Although \({}^{56}\mathrm{Co}\) eventually transitions to stable \({}^{56}\mathrm{Fe}\) and emits additional MeV lines, its contribution only becomes significant at later times due to its relatively long half-life of about 77 days. Consequently, \({}^{56}\mathrm{Co}\) decay can be neglected during the early phase considered here \citep[see, e.g.,][]{1994ApJS...92..527N,2020ApJ...897..152C}. The rest-frame luminosity of the \(0.158\,\)MeV line is (see Methods)
\begin{equation}
L_{\rm 0.158\,MeV}(t) = 7.2 \times 10^{42} \, \left(\frac{M_{\text{Ni}}}{M_\odot}\right) \,  \, e^{-1.31 \times 10^{-6} t} \, \mathrm{erg\,s^{-1}}
\label{eq:Lt}
\end{equation}

The observed luminosity, \(L_{\text{obs}}\), is significantly amplified by relativistic Doppler boosting. 
For a jet moving with a Lorentz factor \(\Gamma\) and velocity \(\beta_{\rm jet} = v_{\rm jet}/c\), the observed energy \(E_{\text{obs}}\) of a photon, originally emitted with a rest-frame energy \(E_{\text{em}}\), is given by:
\begin{equation} \label{eq:Dop}
    E_{\text{obs}} = \frac{E_{\text{em}}}{\Gamma (1 - \beta_{\rm jet} \cos \theta) (1+z)} = \frac{E_{\text{em}} \, \mathcal{D}}{(1+z)},
\end{equation}
\noindent where \(\theta\) is the angle between the jet direction and the observer's line of sight, \textcolor{black}{and $\theta_{\rm j}$ denotes the jet half-opening angle, which constrains
the angular extent of the emitting region}
\citep[]{2017ApJ...850L..24G,2023ApJ...958..126F}, and \(\mathcal{D}\) is the Doppler factor. Since the rest-frame energy \(E_{\text{em}}\) remains constant, the observed energy \(E_{\text{obs}}\) decreases proportionally to \((t_{\rm obs}-t_0)^{-0.70^{+0.05}_{-0.05}}\); see Methods. Recent multidimensional simulations demonstrate that Rayleigh–Taylor instabilities \citep{1978ApJ...219..994C} in supernovae fragment the ejecta into dense \(^{56}\text{Ni}\) clumps or ``bullets'' rather than a smooth flow \citep{2003A&A...408..621K,2010ApJ...714.1371H}, a picture supported by asymmetric iron and nickel line profiles in SN 1987A \citep{1990MNRAS.242..669S,1993ApJ...419..824L,2010ApJ...714.1371H}.  Even if a total \(^{56}\text{Ni}\) mass of 10$^{-3}$ M$_\odot$ were confined into a single spherical bullet at 10$^{16}$ cm (jet half-opening angle 0.02 rad \citep{2024ApJ...962..115R}), its area remains negligible relative to the jet surface.  Thus, individual clumps behave effectively as point sources, yielding a Lorentz-boosted luminosity scaling $\propto \mathcal{D}^4$ \citep{2018pgrb.book.....Z}. Therefore, focusing on the 0.158~MeV line, the observed isotropic luminosity is
\begin{equation}
L_{\text{obs, 0.158 MeV}}(t_{\rm obs}) = 7.2 \times 10^{42} \left(\frac{M_{\text{Ni}}}{M_\odot}\right) e^{-1.31 \times 10^{-6} t_{\rm obs} \mathcal{D}} \, \mathcal{D}^4 \, \mathrm{erg\,s^{-1}},
\label{eq:Lt1}
\end{equation}
where
\[
\frac{M_{\text{Ni}}}{M_\odot} =\mathcal{R} \frac{M_{\text{jet}}}{M_\odot} = \mathcal{R}\frac{\theta_{\rm j}^2 E_{\text{iso}}}{2 \eta\, \Gamma\, M_\odot c^2},
\]
where \(\mathcal{R}\) is the ratio of the mass of Nickel to the mass of the jet,
\(\theta_{\rm j}\) is the half-opening angle of the jet,
\(E_{\text{iso}}\) is the equivalent isotropic energy,
\(\eta\) is the efficiency of the conversion of kinetic energy into radiation,
\textcolor{black}{and in the ultra-relativistic, near-on-axis limit
($\theta \lesssim 1/\Gamma$), corresponding to the special case in which the jet
emission enters the observer’s field of view, the Doppler factor satisfies
$\mathcal{D} \simeq 2\Gamma$},
and \(M_\odot\) is the mass of the Sun.

\subsection{Time-dependent Luminosity}

In the methods section, we demonstrate that the best statistical fits to
the observed luminosity obtained from Zhang et al. (2024)
\citep{2024SCPMA..6789511Z} (Table.~\ref{tab:MeV_Lines}) yield the
$L_{\rm obs} \propto (t_{\rm obs} - 243)^{-1.01^{+0.21}_{-0.26}}$ and for the
energy of the emission line
$E_{\rm line} \propto (t_{\rm obs} - 243)^{-0.70^{+0.05}_{-0.05}}$
{\color{black}; (see {Methods}).
The decay energy, $E_{\text{decay}}$, remains constant during the observation,
from Eq.~\ref{eq:Dop}, the Doppler factor evolves as
$\mathcal{D} \propto (t_{\rm obs} - 243)^{-0.70^{+0.05}_{-0.05}}$.
\textcolor{black}{
Under the standard on-axis approximation $\mathcal{D} \simeq 2\Gamma$, this
implies an effective Lorentz factor evolution
$\Gamma \propto (t_{\rm obs} - 243)^{-0.70^{+0.05}_{-0.05}}$.
}
\textcolor{black}{
Here $\Gamma(t)$ refers to the Lorentz factor of the dominant emitting outflow
component during the mid/late prompt interval in which the line is detected,
and should not be interpreted as the Lorentz factor of the external-shock
afterglow or as the constant $\Gamma$ approximation for a single coasting
prompt shell. Moreover, the inferred Doppler factor evolution should be understood as an effective description of the line-forming region rather than a full dynamical model of the jet.
}

 The luminosity evolution suggests a power-law decay in the observed frame, providing constraints on the jet dynamics and Doppler factor evolution. The injection of nickel from the accretion disk into the jet is proportional to the mass accretion rate during the initial phase, which leads to the mass of nickel being approximated by an increasing power-law relation \(
M_{\text{Ni}} \propto (t_{\rm obs} - 243) ^{\xi},
\) where  \( \xi \) is a constant within the range 1 to 2 \citep{2008MNRAS.388.1729K, 2013ApJ...767L..36W, 2022MNRAS.510.4962G}. This mass evolution leads to the following expression for the observed luminosity:
\begin{equation}
L(t_{\rm obs}) \propto   \mathcal{D}^{4} \, (t_{\rm obs} - 243)^{\xi} \ \mathrm{erg\,s^{-1}}, \rightarrow (t_{\rm obs} - 243)^{-1.01} \propto (t_{\rm obs} - 243)^{-(2.80-\xi)},
\label{eq:Ltt}
\end{equation}
leading to $\xi = 1.79$. To ensure the completeness of the available data, we combined the luminosity measurements from Ravasio et al. (2024) \citep{2024Sci...385..452R} (Table~\ref{tab:Lgauss}) with those from Zhang et al. (2024) \cite{2024SCPMA..6789511Z}. Performing a joint fit, we obtain \( L_{\rm obs} \propto (t_{\rm obs} - 246)^{-1.14^{+0.05}_{-0.04}} \), which corresponds to \( \xi \approx 1.66 \) in line with the theoretical / simulation values \cite{2013ApJ...767L..36W, 2022MNRAS.510.4962G}. Since the best fit for the line energy is identical to that obtained using only Zhang et al. (2024)’s data \cite{2024SCPMA..6789511Z}, we adopted the latter for our analysis, ensuring consistency and reliability. \textcolor{black}{It is worth noting that the model predictions for the jet luminosity, total mass, and entrained $^{56}$Ni mass account exclusively for the 158~keV decay branch. The 24~MeV feature is detected only within a short time interval, preventing a reliable reconstruction of its luminosity evolution. We therefore assume that the temporal evolution of the Doppler factor is identical for both the 12~MeV and 24~MeV components.
}

 To reproduce the isotropic-equivalent luminosity of \(\sim 10^{51}  \mathrm{erg\,s^{-1}}\) observed during 246–256 s, we impose two physical constraints: (i) a maximum isotropic-equivalent kinetic energy of \(7.1 \times 10^{57}  \text{erg}\) \citep{2023ApJ...944L..34R,2023ApJ...950...28R}, and (ii) a jet mass limit \(\lesssim 3 \times 10^{-2}  M_\odot\) to satisfy baryon-loading constraints. These ensure self-consistency in the jet's kinetic energy and Lorentz factor.  In a canonical collapsar framework, the jet's kinetic power originates from the spin energy of a rapidly rotating black hole (BH). For a maximally spinning BH (\(a_* = 1\)) of mass \(M \simeq 11  M_\odot\), the extractable rotational energy is \(
 E_{\rm rot} \simeq \left(1 - \frac{1}{\sqrt{2}}\right) M c^2 \approx 0.293  M c^2 \approx 5.8 \times 10^{54}  \text{erg}.
 \) 
 Assuming a Blandford-Znajek (BZ) efficiency of \(\eta_j \sim 30\%\) for moderate spin (with theoretical extremes up to \(\sim 140\%\) for \(a_* \approx 1\) \cite{2011MNRAS.418L..79T}) and collimation into a half-opening angle \(\theta_{\rm j} \lesssim 0.02  \text{rad}\) (yielding \(\theta_{\rm j}^2/2 \sim 2 \times 10^{-4}\)) \citep{2024ApJ...962..115R,2025arXiv250317765G}, the isotropic-equivalent kinetic energy becomes \(
 E_{\rm kin,iso} = \frac{2  \eta_j  E_{\rm rot}}{\theta_{\rm j}^2} \approx 7.1 \times 10^{57}  \text{erg}.
 \)

\begin{figure}
\centering
\includegraphics[width=0.7\linewidth]{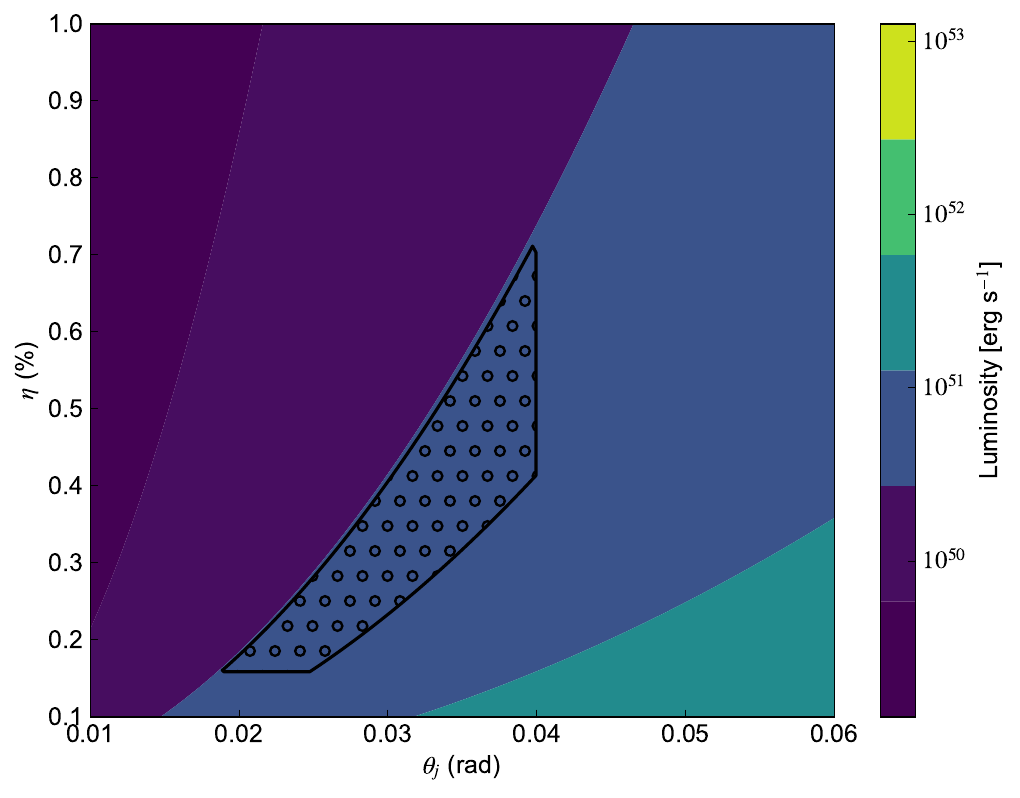} 
\caption{
\textbf{Parameter constraints for the GRB jet-Nickel model}. {The region enclosed by the solid black boundary curve and marked with a dotted fill pattern} shows viable combinations of radiative efficiency $\eta$ and half-opening angle $\theta_{\rm j}$ that satisfy the observed isotropic-equivalent luminosity (246–256\,s) within $1\sigma$ errors, for a nickel mass fraction $\mathcal{R}=40\%$. Our analysis yields $0.2\% \lesssim \eta \lesssim 0.75\%$ and $0.016\,\mathrm{rad} \lesssim \theta_{\rm j} \lesssim 0.04\,\mathrm{rad}$, consistent with: (i) isotropic-equivalent kinetic energies $E_{\rm kinetic,\, iso}=9.8\times10^{56}$–$7.1\times10^{57}\,\rm erg$ \citep{2023ApJ...944L..34R}, (ii) independent jet angle estimates \citep{2024ApJ...962..115R,2025arXiv250317765G}, and (iii) an initial jet mass $<3\times10^{-2}M_\odot$. The upper bound on $\eta$ derives from the $\theta_{\rm j}\lesssim0.04\,\mathrm{rad}$ constraint.}
\label{fig:countur}
\end{figure}

 A BH mass of \(11  M_\odot\) falls within the \(5\)–\(15  M_\odot\) range typical of core-collapse simulations. Earlier work \citep{2023ApJ...944L..34R} argued that \(E_{\rm kin,iso} = 7.1 \times 10^{57}  \text{erg}\) requires \(M > 20  M_\odot\) for \(\theta_{\rm j} \sim 0.061  \text{rad}\), deeming such energies challenging. However, narrower jets (\(\theta_{\rm j} \lesssim 0.02  \text{rad}\)) inferred from recent afterglow analyses \citep{2024ApJ...962..115R,2025arXiv250317765G} resolve this tension by reducing the energy demand by \(\sim 90\%\). Consequently, a rapidly spinning BH of \(M \sim 5\)–\(15  M_\odot\) is compatible with collapsar models under these geometric constraints \citep{2008ApJ...679..639Z}. We adopt a conservative lower mass limit of \(5  M_\odot\), consistent with both core-collapse predictions and energy requirements alleviated by narrow jets.  Under these requirements, our best‐fit Doppler factor and luminosity within 1$\sigma$ error imply a radiative efficiency \(\eta\gtrsim 0.2\%\) and half‐opening angle \(\theta_{\rm j}\gtrsim 0.016\,\mathrm{rad}\), assuming a nickel mass fraction \(\mathcal{R}=40\%\) and \(E_{\rm iso}=1.55\times10^{55}\) erg.  These values agree with independent estimates of \(\theta_{\rm j}\) \cite{2024ApJ...962..115R,2025arXiv250317765G} and \(\eta\) \citep{2023ApJ...944L..34R}. Moreover, to be compatible with the values obtained independently for \(\theta_{\rm j}\), we impose \(\theta_{\rm j}\lesssim0.04\,\mathrm{rad}\) \cite{2024ApJ...962..115R,2025arXiv250317765G} which leads to  \(\eta\lesssim 0.75\%\) (See Fig.~\ref{fig:countur}). From Equation~\ref{eq:Lt1}, for representative values of the jet half-opening angle \(\theta_{\rm j} = 0.03~\mathrm{rad}\) and radiation efficiency \(\eta = 0.2\%\), the entrained nickel mass is estimated as \(\frac{M_{\text{Ni}}}{M_\odot} \approx 5.5 \times 10^{-3}\), and the total jet mass as \(\frac{M_{\text{jet}}}{M_\odot} \approx 1.4 \times 10^{-2}\). These values correspond to kinetic energy corrected for the jet opening angle,  approximately \(3 \times 10^{54}~\mathrm{erg}\).

In fact, Fermi-GBM data suffered severe saturation/pulse-pileup until $\sim\!T_{\rm 0}+ {277}{~\rm second}$, precluding spectral analysis before $\sim\!{280}{~\rm second}$ \citep{2024arXiv241000286B}. Nevertheless, Zhang et al. (2024) \cite{2024SCPMA..6789511Z} detected a line feature at $t\approx{251}{~\rm second}$ using BGO detectors, validated via GECAM-C cross-calibration. While NaI detectors remained saturated, BGO count rates fell below pulse-pileup thresholds by $T_{\rm 0}+{246}{~\rm second}$, enabling reliable spectroscopy in 246–280 s windows. Extrapolating the Doppler-shifted line centroid to $\sim\!{244}{~\rm second}$ with $\theta_{\rm j}={0.03}$ radian and $\eta={0.2}{\%}$ reduces initial nickel mass to $\sim\!7\times10^{-4}M_\odot$. Maintaining $\mathcal{R}={40}{\%}$ permits $\eta \gtrsim {3}{\%}$, correspond to kinetic energy corrected for the jet opening angle,  approximately \(1 \times 10^{53}~\mathrm{erg}\), yielding more conventional jet parameters. This is similar to a collimation-corrected kinetic energy of \(E_{\rm k} \approx 8 \times 10^{52}  \text{erg}\) (Eq. 11, O'Connor et al. 2023 \cite{2023SciA....9I1405O}), consistent with the shallow structured jet model. Moreover, such energetics resolve the ``energy crisis'' identified for top-hat jets (which require \(E_{\rm k} > 4 \times 10^{53}  \text{erg}\)). \textcolor{black}{However, this efficiency crisis can be mitigated by considering coherent effects, which would depend on the Lorentz factor of the shock and the evolution of microphysical \citep[]{2006A&A...458....7I,2013MNRAS.435.3009L}}. In fact, the radiative efficiency of 0.2\%----3\% suggests an inefficient gamma-ray production in mid-late prompt emission as the energy might partition into magnetic fields, protons, or bulk motion \citep[]{2023ApJ...944L..34R,2023ApJ...950...28R}. At the jet‐break epoch in GRB 221009A—observationally identified at $T_{\rm break}\approx1246$ s in X/$\gamma$‑rays by Insight‑HXMT and Fermi/GBM \cite{2024ApJ...962..115R} and at $\sim670$ s in the TeV band by LHAASO \cite{2023Sci...380.1390L}, $\Gamma \lesssim 1/\theta_{\rm j}$ triggers lateral expansion, increasing jet/shock cross-section. This enhances afterglow efficiency to $\sim\!\!10\%$ (vs. prompt $< 10\%$ \citep{2023SciA....9I1405O}) in X‑ray, optical, and TeV bands \citep{2025arXiv250317765G,2024A&A...690A..74G}.

\begin{figure*}%[htbp]
    \centering
    \includegraphics[width=0.620\textwidth]{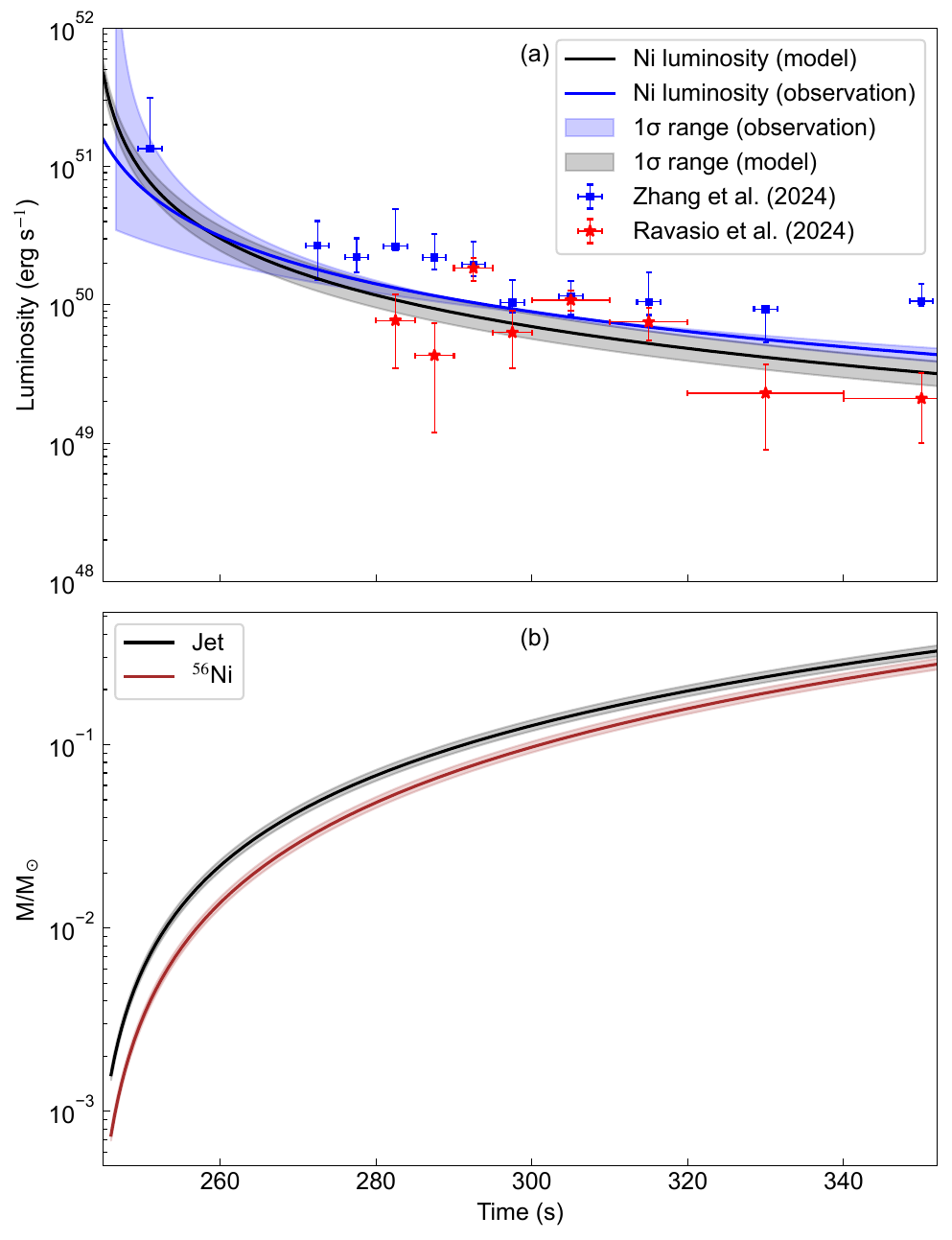}
    \caption{ {\textbf{Line Luminosity Evolution and Inferred Jet–Nickel Mass Growth in GRB 221009A}}. \textbf{(a)}: Red data points represent the luminosity from Ravasio et al. (2024) \citep{2024Sci...385..452R} (Table.~\ref{tab:Lgauss}), while blue points are from Zhang et al. (2024) \citep{2024SCPMA..6789511Z} (Table.~\ref{tab:MeV_Lines}).  {\color{black} Black line shows the l}uminosity evolution corresponding to the best-fit $E_{\rm line}(t)$ model within $1\sigma$ dispersion region defined by $\sigma_{\mathrm{int}}$ (see {Methods}), for a jet opening angle of $\theta_{\rm j} = 0.02$ rad and efficiency $\eta = 0.2\%$. {\color{black} The blue line shows the best-fit joint luminosity evolution, obtained by combining the luminosty datapoints from Ravasio et al. (2024) and Zhang et al. (2024) (see {Time-dependent Luminosity} in Methods).}  \textbf{(b)}: Time evolution of the total jet mass and entrained Nickel mass $M_{\rm Ni}(t) \propto (t_{\rm obs}-243)^{1.6}$, both in solar mass units, within the $1\sigma$ confidence interval derived from the observational best-fit of line's energy vs. time. The initial jet and Nickel masses are approximately $5 \times 10^{-3}\,M_\odot$ and $2 \times 10^{-3}\,M_\odot$, respectively. The total jet mass grows significantly, reaching $\sim 0.1\,M_\odot$ at late times. {\color{black} The  model prediction for the luminosity, the total jet mass and the entrained Nickel mass accounts only for the 158 keV decay branch.}}
    \label{fig:evolution1}
\end{figure*}

For a best-fit model within the $1\sigma$ uncertainty of the spectral line energy evolution $E_{\rm line}(t)$, Fig.~\ref{fig:evolution1} (a) presents the resulting luminosity evolution for a jet half-opening angle $\theta_{\rm j} = 0.024$ rad and an efficiency $\eta = 0.2\%$. As evident from the figure, the predicted luminosity curve lies entirely within the $1\sigma$ confidence interval of the luminosity–time relation obtained from the observational data fit. This agreement supports the consistency of the proposed dynamical scenario with the data. Figure~\ref{fig:evolution1} (b) shows the time evolution of the total jet mass and the entrained nickel mass, $M_{\rm Ni}(t)\propto(t_{\rm obs}-243)^{1.5}$, both in units of solar masses.  Initially, the jet and nickel masses are $\sim5\times10^{-3}M_\odot$ and $\sim2\times10^{-3}\,M_\odot$, respectively, rising to $\sim10^{-1}\,M_\odot$ at later times.  Although such a large jet mass is atypical for gamma-ray bursts, two effects naturally alleviate the need for an excessively massive outflow: First, high-resolution simulations demonstrate that jets confronting heavy baryon loads decelerate and spread laterally much more slowly than in early analytic models.  For a canonical initial opening angle ($\theta_0\approx$ 0.05–0.1), the jet collimation is preserved for hundreds of seconds, with $\theta_{\rm j}\propto t^{0.1-0.2}$ and the bulk Lorentz factor remaining high at $t\sim350\,$s, so that Doppler beaming decays only gently \cite{2012MNRAS.421..570G,2006RPPh...69.2259M}.  A more tightly beamed, slowly spreading jet therefore produces stronger line emission at a given nickel mass—reducing the required $M_{\rm Ni}$ by factors of a few. Second, multi-dimensional explosion models show that nickel is ejected in dense clumps that coast for thousands of seconds without significant mixing \cite{2010ApJ...714.1371H}.  Clumps moving at $\sim4{,}000\, \mathrm{km\,s^{-1}}$ at $t\sim300\,$s retain their inertia until $t\gtrsim10^4\,$s, releasing radioactive energy within the jet core long after naive expansion times.  Together, slow jet spreading and clumped nickel imply that the line-emitting region remains compact and strongly Doppler-boosted over $t\approx250$–350\,s, so that the 0.158MeV line at $t\approx350\,$s can be produced with much less nickel than a simple spherical estimate would suggest.

Finally, instrumental systematics in the brightest phases of GRB221009A introduce a further potential reduction in the inferred nickel mass.  Fermi/GBM suffered dead‐time and pulse‐pileup during the prompt peak, which Zhang et al. used GECAM-C’s unsaturated data to correct, finding consistent cross‐normalization once the worst intervals passed \cite{2024SCPMA..6789511Z}.  Residual uncertainties in GECAM’s high-gain channel, however, imply the earliest line fluxes may be overestimated by $\sim15$–25\%.  Although the feature remains significant at $>5\sigma$ and shows no instrumental origin \cite{2024arXiv241000286B}, these systematics—and the luminosity discrepancy between Ravasio et al. and Zhang et al. at late times—suggest a conservative one order of magnitude uncertainty in late fluxes.  A modest reduction in the measured flux correspondingly lowers the nickel mass required to match the observed line luminosity.

{\color{black}\subsection{Nickel Mass in the Ultrarelativistic Jet vs Nickel Synthesized in the Supernova Ejecta}

Independent light curve modeling of GRB 221009A/SN 2022xiw \citep[e.g.][]{2023ApJ...949L..39S,2024ApJ...971...56K} shows ejecta masses of several $M_{\odot}$ and nickel masses up to $0.25~M_{\odot}$, with an upper limit of $0.36~M_{\odot}$, indicating that the JWST-inferred $0.09~M_{\odot}$ \citep{blanchard2024jwst} is not a strict upper bound. In addition, collapsar simulations indicate that vigorous nucleosynthesis in the accretion disk \citep{MacFadyen:1998vz,2006ApJ...647.1255N}, together with mixing and entrainment through disk winds, can enrich the polar regions with nickel. This, together with the extreme nature of GRB 221009A, provide a physically plausible channel for locally nickel-rich jets without requiring a globally large nickel fraction in the full ejecta. Therefore, a comparison between our inferred nickel mass in the ultrarelativistic jet and the late-time estimate of JWST  ($\approx0.09~M_{\odot}$) should be made with some caution. The JWST value represents the {total} nickel synthesized in the supernova ejecta and observed $\sim$ 170 days after the explosion, while our MeV line analysis reflects only the fraction of nickel entrained in the ultrarelativistic jet during the first few hundred seconds. 

}

 \section*{Conclusions}

In summary, \textcolor{black}{
we show that the MeV emission line detected in GRB~221009A during the prompt phase,
which exhibits a coherent temporal and spectral evolution with its centroid
energy shifting from $\sim$37 to $\sim$6~MeV,}
is compatible with Doppler-boosted radioactive decay of freshly synthesized
$^{56}$Ni (rest-frame 158~keV) entrained in the relativistic outflow.
\textcolor{black}{
In addition, we find tentative evidence for a higher-energy excess near $\sim$24~MeV
at 290--300~s, consistent with the boosted 270~keV decay branch and supportive of
this interpretation, but not required for the main conclusion.} While the apparent temporal evolution of the line centroid  and luminosity is broadly consistent with expectations from radioactive-decay kinematics coupled to standard jet-dynamics models, we emphasize that these power-law indices are subject to significant uncertainties. In particular, instrumental pile-up and background subtraction challenges may bias both slopes and normalizations. Obtaining cleaner, higher-resolution spectra—ideally from bursts of similar brightness observed with minimal pile-up—will be essential to refine these evolutionary trends and pin down their true physical parameters.  \textcolor{black}{
We stress that the absence or marginal detection of higher-energy decay branches
does not refute the proposed scenario, as energy-dependent opacity, jet geometry,
and instrumental sensitivity naturally suppress such features in the observed
spectrum.
}

 {\color{black} It is important to note that \textcolor{black}{our analysis of the MeV spectral features} in GRB 221009A indicates a broader context of nickel production in GRB-SN systems. The JWST results primarily trace nickel in the {slow-moving} ejecta powering the optical/NIR light curve, while our prompt line detection isolates the nickel component in the {fast-moving} jet. Therefore, these two measurements should be considered complementary, probing distinct physical reservoirs of nickel. Although our inferred jet nickel mass is indeed high at later time, it may reflect the extraordinary conditions of this ``brightest-of-all-time'' event, highlighting the need for future simulations of extreme collapsar systems.}

These findings  establish {a} spectroscopic link between prompt GRB emission and supernova nucleosynthesis, which opens {an additional}  observational window into collapsar jets. The identification of Lorentz-boosted nuclear lines calls for a re-evaluation of jet composition, baryon loading, and energy-extraction mechanisms in long GRBs, and highlights the role that next-generation high-resolution $\gamma$-ray spectrometers must play in future missions. Continued multi-wavelength and polarimetric follow-up of exceptionally bright bursts will be vital to map the interplay between heavy-element synthesis and jet dynamics across the full GRB population.

\textcolor{black}{We finally note that, while the present framework captures the essential kinematic and radiative ingredients required to explain the observed MeV features, a complete treatment will require time-dependent multidimensional radiative-transfer simulations that self-consistently couple jet dynamics, nuclear decay, and photon propagation.}

\section*{Methods}

\subsection{Spectral Analysis}

\textcolor{black}{We follow standard procedures for gamma-ray spectral analysis using the software packages \texttt{rmfit} (version 4.3.2) \citep{rmfit_432} and \texttt{threeML} (version 2.4.2) \citep{threeML_docs}, and independently verify the results by \texttt{XSPEC} (version 12.15.1) \citep{xspec_12151} with spectra exported by \texttt{GBM Data Tools} (version 1.1.1)\citep{GbmDataTools}. Time-tagged event (TTE) data from the Fermi Gamma-ray Burst Monitor (GBM; NaI7 and BGO1 detectors) are extracted for the time interval between 290 and 300~s after the trigger. The background is modeled from pre- and post-burst intervals using polynomial fits in count space and subtracted before spectral fitting.}

\textcolor{black}{Spectra are jointly fitted in \texttt{rmfit} using the Castor C-statistic as the likelihood function, consistent with Poisson counting statistics. The same spectral intervals are analyzed with \texttt{threeML}, which allows for fully Bayesian inference through Markov Chain Monte Carlo (MCMC) sampling. The posterior distributions of the spectral parameters and line fluxes are verified to be unimodal and converged, with Gelman–Rubin statistics $R<1.05$ for all parameters.}

\textcolor{black}{From the spectral fitting of the observed emission between 290 and 300s ({See} Fig. \ref{fig:270line}), the parameters of the Band function are determined to be $A = 0.318 \pm 0.002$, $E_{\mathrm{p}} = 1014 \pm 58.1~\mathrm{keV}$, $\alpha = -1.685 \pm 0.004$, and $\beta = -2.491 \pm 0.104$. Two additional Gaussian components are required to model the emission features, centered at $E_{1,\mathrm{obs}} = 12.22 \pm 0.03~\mathrm{MeV}$ and $E_{2,\mathrm{obs}} = 24.24 \pm 0.12~\mathrm{MeV}$, respectively.}

\subsection{Second Line Significance}
The addition of a second Gaussian \textcolor{black}{component}  to the Band + first-line  reduces the Poisson log-likelihood (Castor C-STAT) from $C_{0}=830.07$ to $C_{1}=821.14$ at time 290-300 s, giving a likelihood-ratio test statistic $\Delta C = C_{0}-C_{1}=8.93$.

If the three new parameters (amplitude $ A_{2} $, centroid $ E_{2} $, and width $ \sigma_{2} $, can vary freely on the real line, the regularity conditions \cite{wilks1938large} apply and
\begin{equation}
\Delta C \xrightarrow[]{N\rightarrow\infty} \chi^{2}_{3}.
\end{equation}
The tail probability is
\begin{equation}
p_{\chi^{2}_{3}} = 1 - F_{\chi^{2}_{3}}(8.93) = 0.0302.
\end{equation}
This corresponds to a one-sided Gaussian significance
\begin{equation}
Z_{\chi^{2}_{3}} = 1.88\;\sigma,
\quad\text{or a confidence level}\quad \mathrm{CL}_{\chi^{2}_{3}} = 96.98\%.
\end{equation}

When the null hypothesis sets $ A_{2}=0 $, the parameter space boundary is reached and the asymptotic distribution follows \cite{chernoff1954distribution}
\begin{equation}
\Delta C \xrightarrow[]{N\rightarrow\infty} \tfrac12\delta(0)+\tfrac12\chi^{2}_{1}.
\end{equation}
Using this mixture,
\begin{equation}
p_{\text{mix}}=\tfrac12\!\left[1-F_{\chi^{2}_{1}}(8.93)\right]=1.40\times10^{-3},
\end{equation}
which implies
\begin{equation}
Z_{\text{mix}} = 2.99\;\sigma,
\quad\text{or a confidence level}\quad \mathrm{CL}_{\text{mix}} = 99.86\%.
\end{equation}

The two analytic limits bracket the true significance
$ 1.9\,\sigma \le Z \le 3.0\,\sigma$. Because the centroid and width are undefined when $ A_{2}=0 $, the mixture distribution is more appropriate than the regular $ \chi^{2}_{3} $ law \cite{2002ApJ...571..545P}. This significance range is consistent with the line fitting significance of $2.5\, \sigma$ that computed by the fitted line flux ($0.038~\mathrm{photon~cm}^{-2}~\mathrm{s}^{-1}$) divided by its uncertainty ($0.015 ~\mathrm{photon~cm}^{-2}~\mathrm{s}^{-1}$).

{\color{black}\subsection{MCMC Analysis as Additional Evidence for the Second Line}

The spectrum of 290–300~s is fitted using a Markov Chain Monte Carlo (MCMC) approach to obtain the full posterior distributions of the model parameters. Each fit used 50 independent chains, with 20\,000 iterations per chain and a burn-in of 2\,000 steps. The posterior samples from these runs are used to calculate the Deviance Information Criterion (DIC)\cite{spiegelhalter2014deviance}, the Widely Applicable Information Criterion (WAIC)\cite{watanabe2013widely}, and the corresponding approximate Bayes factors for quantitative model comparison \cite{kass1995bayes}.

The DIC is defined as
\begin{equation}
\mathrm{DIC} = 2\,\overline{D(\theta)} - D(\bar{\theta}),
\end{equation}
where \( D(\theta) = -2\ln p(D|\theta) \) is the deviance, 
\( \overline{D(\theta)} \) is its posterior mean, and \( D(\bar{\theta}) \) is the deviance evaluated at the posterior mean parameter values. DIC penalizes overfitting by including an effective number of parameters, but it assumes that the posterior is approximately Gaussian and unimodal. This assumption can fail when parameters lie near boundaries or when the posterior is asymmetric, as occurs when the amplitude of a Gaussian \textcolor{black}{component}  approaches zero.

The WAIC is a more general and fully Bayesian measure of predictive performance. It is defined as
\begin{equation}
\mathrm{WAIC} = -2\,\mathrm{elpd}_{\mathrm{WAIC}},
\end{equation}
where the expected log pointwise predictive density ($\mathrm{elpd}$) is
\begin{equation}
\mathrm{elpd}_{\mathrm{WAIC}} =
\sum_{i=1}^{N}
\left[
\ln\!\left(\frac{1}{S}\sum_{s=1}^{S} p(y_i|\theta^{(s)})\right)
- V_s\!\left(\ln p(y_i|\theta^{(s)})\right)
\right].
\end{equation}
Here \( N \) is the number of spectral bins, \( S \) is the number of MCMC samples, and \( V_s \) denotes the variance over the posterior draws. WAIC uses the full posterior distribution to estimate how well the model predicts unseen data and remains valid for non-Gaussian, multimodal, or boundary-constrained posteriors. For this reason, WAIC is more suitable than DIC in our analysis. The Gaussian \textcolor{black}{component}  amplitude in these models can be close to zero, producing asymmetric or truncated posterior distributions that violate DIC’s assumptions.

The model comparison results are summarized in Table~\ref{tab:model_comparison}. Both DIC and WAIC are lower (better) for the two-line model (Band + G + G) compared to the one-line model (Band + G). Specifically, we find $\Delta \text{DIC} \approx 4.34$ and $\Delta \text{WAIC} \approx 4.56$ in favor of the two-line model. This corresponds to a Bayes factor of roughly 
\begin{equation}
\mathrm{BF}_{(\mathrm{Band+G+G,\,Band+G})} \approx
\exp\!\left(-\tfrac{1}{2}\Delta\mathrm{WAIC}\right) \approx 9.8.
\end{equation}
In other words, given the data, the model with two lines is about ten times more likely than the model with one line. In qualitative terms, \textcolor{black}{this provides positive but moderate evidence for an additional spectral component during the 290–300 s interval.} We caution that this is ``moderate'' rather than ``strong'' evidence in Bayesian terms, but it bolsters the $\sim 3\sigma$ frequentist result with an independent measure that naturally accounts for the look-elsewhere effect.

Regarding the concern that the line was found in a specific $10$~s slice (out of a longer burst): the Bayesian approach inherently conditions on having chosen that slice by evaluating the evidence within that slice. If we had blindly searched many intervals, we would need to lower the significance, but here the interval selection was guided by prior analysis that identified the $\sim 12$~MeV line region. We also tested adjacent intervals (e.g., $280-290$~s, $300-310$~s) and confirmed that the two-line model is not favored outside $290-300$~s. Thus, the Bayes factor we compute is already reflective of the targeted search and does not require an additional trials penalty. Essentially, the analysis answers: ``Given a hint of an excess in $290-300$~s, how much more likely is it that two lines are present as opposed to one?'' and finds about an order of magnitude preference for two lines.

We note that our spectrum (Figure~\ref{fig:270line}) looks somewhat different from that in Ravasio et al. (2024) \cite{2024Sci...385..452R}. Their broader $280-320$~s integration dilutes the second line and results in only one obvious hump $\sim 12$~MeV. Our finer time selection isolates the moment when the second line briefly emerged. The MCMC analysis confirms that within $290-300$~s, the second line is statistically warranted. Differences between our spectrum and previous papers are primarily due to this narrower time window and possibly slight differences in data reduction (we applied standard Fermi/GBM analysis procedures consistent with Ravasio et al. (2024) \cite{2024Sci...385..452R}, so any residual differences likely stem from the time selection and the inclusion of GECAM-C data in Zhang et al \cite{2024SCPMA..6789511Z}).

In conclusion, the Bayesian model selection indicates that the two-line model is moderately favored over the single-line model. This quantitative result supports the presence of the second line during the 290-300 s interval.

}

\subsection{\textcolor{black}{$\gamma\gamma$ Annihilation of Nuclear Decay Lines by Prompt Photons}}

For a relativistic shell with Doppler factor $\mathcal{D}$ at redshift $z$, the observed photon energy $E_{\mathrm{obs}}$ is converted to the co-moving frame

\begin{equation}
E'=\frac{E_{\mathrm{obs}}\,(1+z)}{\mathcal{D}}.
\label{eq:comoving-energy}
\end{equation}

\textcolor{black}{Throughout this work, quantities denoted by a prime correspond to measurements
in the comoving frame of the outflow. When additional clarification is required,
we explicitly indicate the reference frame using subscripts (e.g., observer-frame
quantities are labeled with the subscript ``obs'').}

Band photon spectrum is observed, as a component of the entire spectrum. The differential photon flux (photons s$^{-1}$ cm$^{-2}$ keV$^{-1}$) is

\begin{equation}
\frac{\mathrm{d}N}{\mathrm{d}E}
   =
   \begin{cases}
   A\!\left(\dfrac{E}{100~\mathrm{keV}}\right)^{\alpha}
      \exp\!\left(-\dfrac{E}{E_{0}}\right),
      & E < E_{\mathrm b},\\[8pt]
   A\!\left(\dfrac{(\alpha-\beta)E_{0}}{100~\mathrm{keV}}\right)
     ^{\alpha-\beta}
      \exp(\beta-\alpha)
      \left(\dfrac{E}{100~\mathrm{keV}}\right)^{\beta},
      & E \ge E_{\mathrm b},
   \end{cases}
\label{eq:band-spectrum}
\end{equation}
where $E_{0}=E_{\mathrm{p}}/(2+\alpha)$ and $E_{\mathrm b}=(\alpha-\beta)\,E_{0}$ \citep{1993ApJ...413..281B}.

Using photon number conservation and relativistic conversion, the differential comoving number density (photons cm$^{-3}$ keV$^{-1}$) at radius $R$ is \cite{2011ApJ...726...89Z}

\begin{equation}
n'(E';R)=
\frac{D_{L}^{2}}{R^{2}\,c\,(1+z)^2}
\,
\left.\frac{\mathrm{d}N}{\mathrm{d}E}\right|_{E=E_{\mathrm{obs}}=\mathcal{D} E' /(1+z)},
\label{eq:comoving-density}
\end{equation}
{\color{black}where $D_{L}$ is the luminosity distance.}

The angle averaged Breit-Wheeler pair production cross section gives \citep{2010PhR...487....1R} 
\begin{equation}
\bar{\sigma}_{\gamma\gamma}(s) =
\begin{cases}
\displaystyle
\frac{3\,\sigma_{\mathrm{T}}}{16}\,(1-\mu^{2})
\left[
(3-\mu^{4})\ln\left(\frac{1+\mu}{1-\mu}\right)
-2\mu\,(2-\mu^{2})
\right],
& s\ge1,\\[8pt]
0, & s<1.
\end{cases}
\label{eq:sigma_gg}
\end{equation}
where  $\mu = \sqrt{1-1/s} $, and $\sigma_\mathrm{T}$ is the Thomson cross section, {\color{black} and $s$ is the dimentionless centre-of-mass energy squared of the two photons of energies $E'_1$ and $E'_2$ 
\begin{equation}
s=\frac{E'_1\,E'_2}{(m_{\mathrm e} c^{2})^{2}}.
\label{4}
\end{equation}}

In the comoving frame, for a test photon of energy $E'_1$, the \textcolor{black}{opacity} is
\begin{equation}
\kappa_{\gamma\gamma}(E'_1;R) =
\int_{E'_{\min}}^{\infty}
n'(E'_2;R)\;
\bar{\sigma}_{\gamma\gamma}\!\left(\frac{E'_1\,E'_2}{(m_{\mathrm{e}}c^{2})^{2}}\right)
\,\mathrm{d}E'_2.
\label{eq:kappa}
\end{equation}
where
\begin{equation}
E'_{\min} = \frac{(m_{\mathrm{e}}c^{2})^{2}}{E'_1}\ ,
\label{eq:Emin}
\end{equation}
is the minimal photon energy by which the total center-of-mass energy is enough to produce the rest mass of the electron and positron.

Assuming the path length is approximately the radial distance $R$ where the emission occurs, the optical depth gives
\begin{equation}
\tau_{\gamma\gamma}(E'_1;R)=
\frac{R}{\Gamma}\,\kappa_{\gamma\gamma}(E'_1;R).
\label{eq:optical-depth}
\end{equation}

The attenuated photon flux of the Gaussian \textcolor{black}{components}  is then by multiplication with an attenuation factor, which indicates the survival probability
\begin{equation}
e^{-\tau_{\gamma\gamma}}.
\label{eq:attenuation-factor}
\end{equation}

Because $n' \propto R^{-2}$ from equation~\ref{eq:comoving-density}, equation~\ref{eq:kappa} indicates that $\kappa_{\gamma\gamma} \propto R^{-2}$, and equation~\ref{eq:optical-depth} then gives $\tau_{\gamma\gamma} \propto R^{-1}$. Therefore, the attenuation is more severe when the annihilation occurs at smaller radii.

From the spectral fitting of the observed emission between 290 and 300 seconds ( Fig.~\ref{fig:270line}), the parameters of the Band function are determined to be $A = 0.318$, $E_{\mathrm{p}} = 1014\,\mathrm{keV}$, $\alpha = -1.685$, and $\beta = -2.491$. The Band spectrum is attributed to internal shocks occurring at a radius $R$ of order $10^{12}$–$10^{15}\,\mathrm{cm}$ \cite{2024arXiv240816748V}. Gaussian emissions arising from radioactive decay interact with the Band spectrum. Using the equations described above, we computed the optical depth (equation~\ref{eq:optical-depth}), as shown in Fig.~\ref{fig:optical-depth} (a). Higher energy emission lines are associated with larger collision cross-sections. If the prompt emission takes place at a radius of $\sim 10^{14}\,\mathrm{cm}$, the highest energy line at $1562\,\mathrm{keV}$ is expected to be strongly suppressed (attenuation factor $\sim 0.007$), while the lower-energy lines at $158\,\mathrm{keV}$ and $270\,\mathrm{keV}$ are less affected (attenuation factor $\sim 0.8$ and $0.7$, respectively). If the emission occurs at $\sim 10^{13}\,\mathrm{cm}$ or smaller radii, all lines will be significantly suppressed. 

 We also notice that between 250 and 300 seconds, the count rate decreases approximately as $t^{-1.5}$ \cite{2023ApJ...949L...7F}, and the Lorentz factor $\Gamma$ decreases approximately as $t^{-0.7}$ (Fig. \ref{fig:mcmcfit3}). The collision radius of the internal shock $R$ scales as $\Gamma^2$, so the optical depth scales ($\propto \frac{A}{R~ \Gamma}$) approximately as $t^{0.6}$. If the variability of the central engine remains constant, the optical depth will increase by time, hence even the lines of lowest energies will disappear from observation due to attenuation at later times.

 The final results are shown in Fig.~\ref{fig:nickel-decay-lines}, where we assume that the internal shock occurs at $10^{14}\,\mathrm{cm}$. Although the original energy flux of the $1562\,\mathrm{keV}$ line is in the same order of magnitude of other lines, it is annihilated to be $< 1\%$ and thus not distinguishable by Fermi-LAT because it is much less than the inverse Compton flux. The $158\,\mathrm{keV}$ and ${\color{black}270}\,\mathrm{keV}$ emission lines are both annihilated by tens of percent, but the $158\,\mathrm{keV}$ line is less affected. These two lines fall within the observation range of Fermi-GBM and can be detected, as is indeed the case.

\begin{figure}[h]
    \centering
    \includegraphics[width=0.6\textwidth]{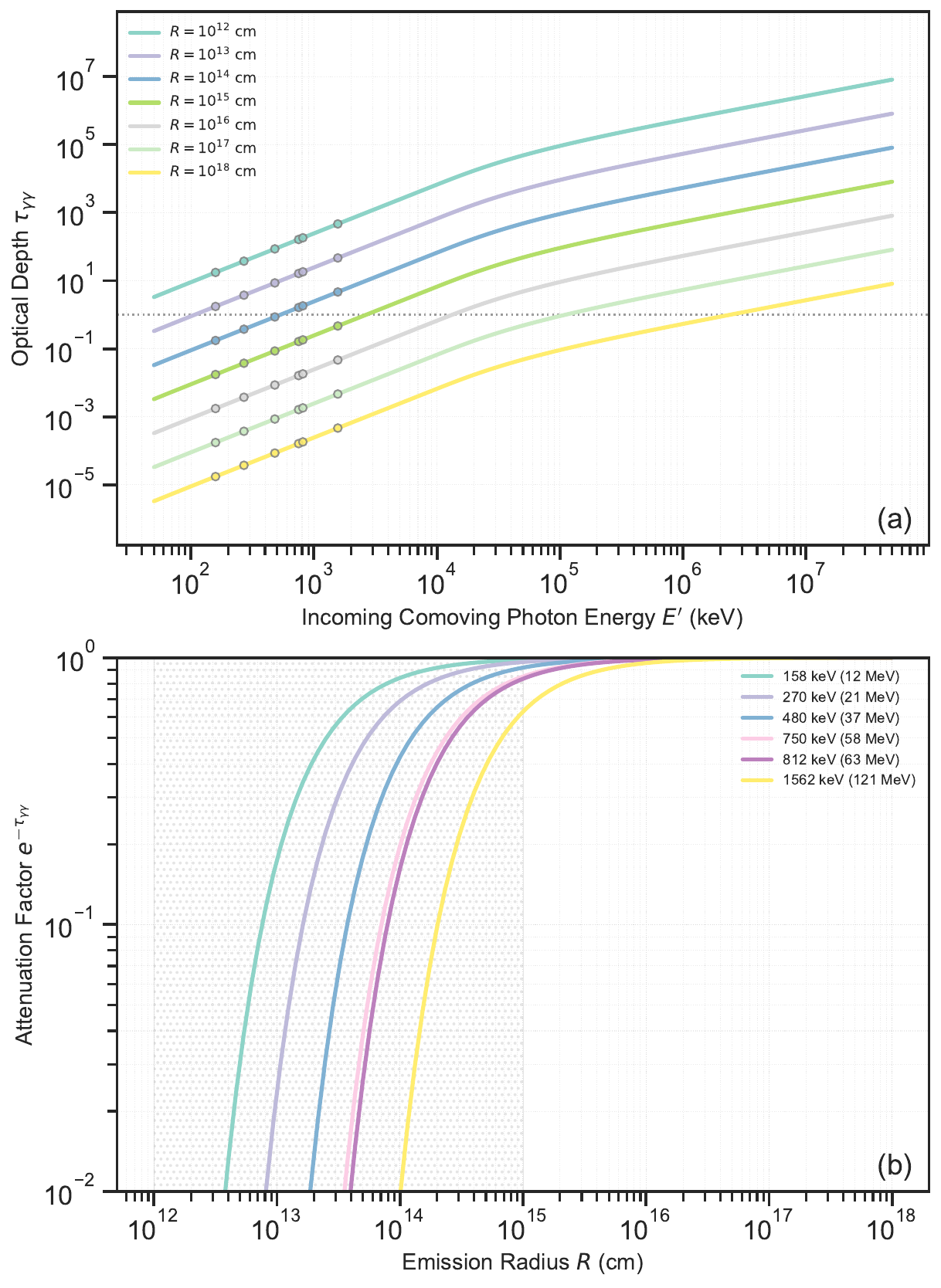}
    \caption{\textbf{Radius-Dependent $\gamma\gamma$ Opacity and Survival of Doppler-Boosted Nickel Lines}. (a): Gamma-gamma annihilation optical depth within the GRB jet  at time 290-300 s. Calculated optical depth, $\tau_{\gamma\gamma}$, for radioactive photons traversing the prompt emission field. Curves represent different assumed emission radii ($R$) from the central engine, ranging from $10^{12}$ cm to $10^{18}$ cm. The target photon field is modelled using the observed Band function component of the GRB spectrum transformed into the comoving frame of energy $E'$. Circles mark the calculated optical depths for the comoving frame centre energies of the $^{56}\mathrm{Ni}$ decay lines. The dashed horizontal line indicates $\tau_{\gamma\gamma}=1$. (b): Attenuation factor for $^{56}\mathrm{Ni}$ lines due to $\gamma\gamma$ annihilation at time 290-300 s. The plot shows the attenuation factor (survival probability $e^{-\tau_{\gamma\gamma}}$) for photons originating from specific $^{56}\mathrm{Ni}$ decay lines identified by their comoving energy (boosted energy in the parenthesis) as a function of the emission radius ($R$) within the jet. The optical depth $\tau_{\gamma\gamma}$ for each line's comoving energy is calculated using the comoving Band spectrum as the target field (see Methods and panel (a)). Higher energy lines experience significant attenuation at smaller radii.}
    \label{fig:optical-depth}
\end{figure}

\subsection{The rest-frame luminosity of the 0.158 MeV line}

\noindent Radioactive \(^{56}\text{Ni}\) undergoes electron capture decay to form \(^{56}\text{Co}\), as described by \citep{1994ApJS...92..527N}:
\[
^{56}\mathrm{Ni} \to ^{56}\mathrm{Co} + \gamma + \nu_e.
\]
The half-life of \(^{56}\text{Ni}\) is 6.10 days, corresponding to a mean lifetime of 8.80 days \citep{1994ApJS...92..527N}. The decay process is accompanied by the emission of $\gamma$-ray photons. Table~\ref{tab:ni_decay} summarizes the key gamma transitions associated with \(^{56}\text{Ni}\) decay. For a relativistic jet associated with a GRB, these gamma-rays are Doppler-boosted, potentially becoming observable as high-energy line profiles.

\noindent To estimate the luminosity produced by the decay of \(^{56}\text{Ni}\) to \(^{56}\text{Co}\) in a supernova remnant containing  \(^{56}\text{Ni}\), we account for the energy released per decay and the temporal evolution of the decay process. Given the molar mass of \(^{56}\text{Ni}\) (\(56 \, \text{g/mol}\)) and Avogadro's number (\(6.022 \times 10^{23} \, \text{atoms/mol}\)), the total number of \(^{56}\text{Ni}\) atoms is:  
\begin{equation}
N_0 = \frac{M_{\text{Ni}}}{56 \, \text{g/mol}} \times 6.022 \times 10^{23} \approx 2.14 \times 10^{55} \, \frac{M_{\text{Ni}}}{M_\odot}\,\text{atoms} .
\end{equation}
where $M_{\text{Ni}}$ is the mass of Nickel. The decay of \(^{56}\text{Ni}\) has a half-life of \(t_{1/2} = 6.10 \, \text{days}\approx 5.27 \times 10^5 \, \text{s}\), which leads the decay constant \(\lambda\) given by:  
\begin{equation}
\lambda = \frac{\ln(2)}{t_{1/2}} = \frac{0.693}{5.27 \times 10^5} \approx 1.31 \times 10^{-6} \, \text{s}^{-1}.
\end{equation}

\noindent The instantaneous luminosity due to radioactive decay at time \(t\) is:  
\begin{equation}
L(t) = N_0 \, \lambda \, Q_{\gamma} \, e^{-\lambda t}.
\end{equation}
which is \(Q_{\gamma} = 1.75\, \rm MeV\) the total energy emitted via gamma photons \cite{1994ApJS...92..527N}. Substituting the values:  
\begin{equation}
L(t) = 7.86 \times 10^{43} \, \left(\frac{M_{\text{Ni}}}{M_\odot}\right) \,  \, e^{-1.31 \times 10^{-6} t} \, \mathrm{erg\,s^{-1}}
\label{eq:Lt-1}
\end{equation}
where \(t\) is measured in seconds in rest-frame time of the source. The total gamma-ray luminosity (\(L_{\text{total}}\)) emitted during the decay of \(^{56}\mathrm{Ni}\) to \(^{56}\mathrm{Co}\) is calculated by summing the contributions from all de-excitation pathways, where each transition’s luminosity is proportional to its photon energy (\(E_i\)) multiplied by its branching intensity (\(I_i\)), $L_{\text{total}} = \sum_i E_i \, I_i$ , as listed in Table.~\ref{tab:ni_decay}. 
yielding \(L_{\text{total}} \approx 1.718 \, \text{MeV/decay}\). The \(0.158\,\)MeV line (\(3^+ \rightarrow 4^+\)), despite its near-unity branching ratio (\(98.8\%\)), contributes only \(9.1\%\) of \(L_{\text{total}}\) due to its lower energy relative to higher-energy transitions, with the dominant contributor being the \(0.812\,\)MeV line (\(2^+ \rightarrow 3^+\)) at \(40.6\%\). Therefore, the rest-frame luminosity of the \(0.158\,\)MeV line is

\begin{equation}
L_{\rm 0.158\,MeV}(t) = 7.2 \times 10^{42} \, \left(\frac{M_{\text{Ni}}}{M_\odot}\right) \,  \, e^{-1.31 \times 10^{-6} t} \, \mathrm{erg\,s^{-1}}
\label{eq:Lt}
\end{equation}

\subsection{Best fit procedure}

\noindent We modeled the log–log relation between line energy and luminosity with a Bayesian linear regression including intrinsic scatter. The model is defined as $
y = \alpha + m x, $ where \(x = \log_{10}(\text{energy})\) and \(y = \log_{10}(\text{luminosity})\).  The log-likelihood function is then formulated as:

\[
\ln \mathcal{L} = -\frac{1}{2} \sum_{i} \left[ \ln \left(2\pi\, \sigma_{\text{tot}, i}^2\right) + \frac{\Delta_i^2}{\sigma_{\text{tot}, i}^2} \right].
\]

\noindent Where for each datum, the total uncertainty is given by 
\( \sigma_{\text{tot}} = \sqrt{\sigma_{\text{meas}}^2 + \sigma_{\text{int}}^2},\). To account for additional scatter not captured by the measurement errors, an intrinsic dispersion term, \(\sigma_{\text{int}}\), was incorporated \citep{dagostini2005}. \(\sigma_{\text{meas}}\) is selected asymmetrically based on the residual \(\Delta = y - (\alpha + m x)\) (using the lower error for negative \(\Delta\) and the upper error otherwise). Uniform priors were imposed on the parameters, restricting \(\alpha \in (-50, 50)\), \(m \in (-10, 10)\), and \(\sigma_{\text{int}} \in (0, 1)\). The parameter space was explored using an affine-invariant ensemble MCMC sampler (emcee), and the medians of the resulting posterior distributions were adopted as the best-fit estimates.

\begin{figure*}[ht!]
\centering
\includegraphics[width=\linewidth]{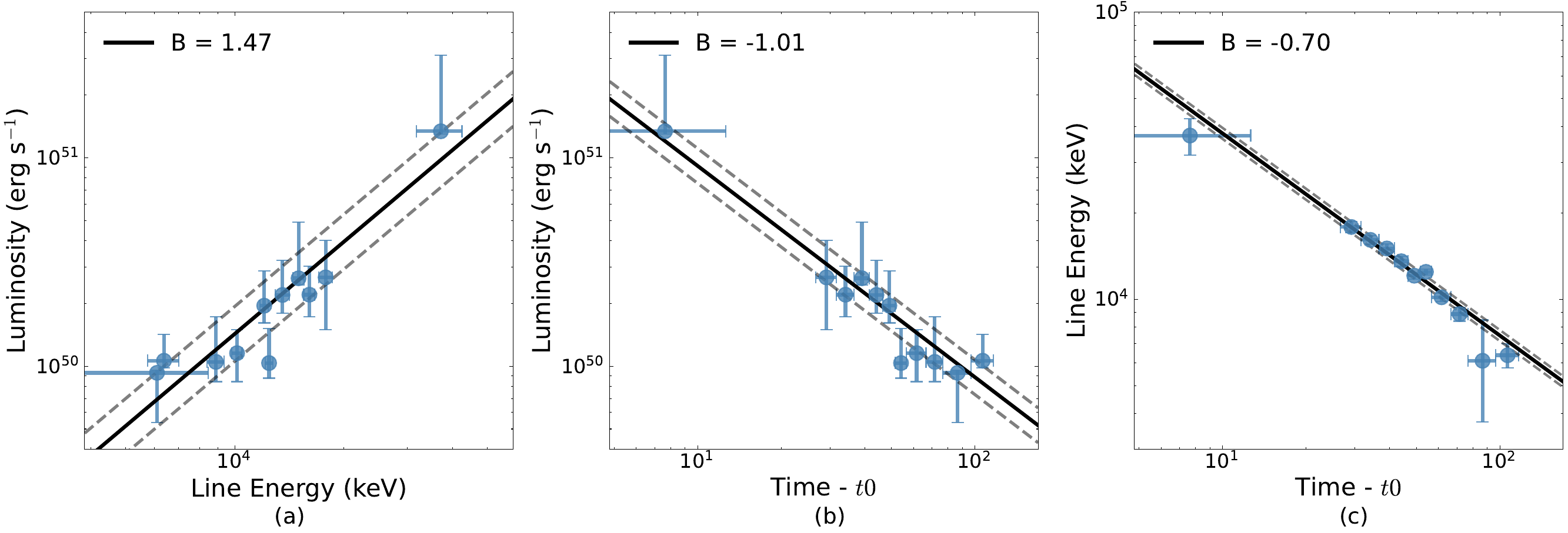} 
\caption{ {\color{black} Spectral–temporal correlations of the MeV line components in GRB 221009A observed by GECAM-C and \emph{Fermi}/GBM. Blue points show the measured values from {taken} from \cite{2024SCPMA..6789511Z}. {The error bars correspond to 1$\sigma$ errors}.  The solid black lines represent the best-fit power-law relations}.  (a) shows the luminosity of the emission line ($L_{\rm obs}$) as a function of the central energy of the emission line ($E_{\rm line}$) {\color{black} fitted with  $L_{\rm obs} = A (E_{\rm line})^B$ with the medians of the resultant posterior distribution yielding $\log(A) = 44.3^{+0.96}_{-0.90}$ and $B = 1.47^{+0.21}_{-0.23} $.} (b) displays the luminosity of the emission line ($L_{\rm obs}$) as a function of time {\color{black} fitted with  $L_{\rm obs} = A (t - t_0)^B$ giving $\log(A) = 51.98^{+0.50
}_{-0.39}$, $t_0 = 243^{+4.05}_{-6.00}$ s and $B = -1.01^{+0.21}_{-0.26}$}, while  (c) illustrates the central energy of the emission line ($E_{\rm line}$) as a function of time, {\color{black} fitted with $E_{\rm line} = A (t - t_0)^B$ with $\log(A) = 5.29^{+0.09}_{-0.09}$ and $B = -0.70^{+0.05}_{-0.05}$ ( for fixed $t_0 = 243$\textcolor{black}{s})}. The dashed lines show the 1 $\sigma$ dispersion envelope defined by $\sigma_{\mathrm{int}}$. 
}
\label{fig:mcmcfit3}
\end{figure*}

\noindent The objective is to fit the evolution of $L_{\rm obs}$ vs $E_{\rm line}$,  $L_{\rm obs}(t)$, and $E_{\rm line}(t)$. In the case of $L_{\rm obs}(t)$ and $E_{\rm line}(t)$, we fit assuming a power-law function of the form 
$A (t - t_0)^B$. To establish a robust statistical determination of $t_0$, we first fit $L_{\rm obs}$ as a function of $E_{\rm line}$, expressed as $L_{\rm obs} = A (E_{\rm line})^B$, using the above method. The medians of the resultant posterior distribution yielded a result of $\log(A) = 44.3^{+0.96}_{-0.90}$ and $B = 1.47^{+0.21}_{-0.23} $. Only errors in luminosity were accounted for, as the errors in time were negligible.

\noindent We then fit for $L_{\rm obs}(t)$, finding $\log(A) = 51.98^{+0.50
}_{-0.39}$, $t_0 = 243^{+4.05}_{-6.00}$ s and $B = -1.01^{+0.21}_{-0.26}$. Fixing $t_0 = 243$ \textcolor{black}{s} and repeating the fit for $E_{\rm line}(t)$, we obtain $\log(A) = 5.29^{+0.09}_{-0.09}$ and $B = -0.70^{+0.05}_{-0.05}$ ({\color{black} see Figure \ref{fig:mcmcfit3}})

%$E_{\rm line} \propto (t - 243)^{-0.70}$.

\noindent Based on these findings, we adopt the best-fit model:
\begin{equation}
    L_{\rm obs} \propto (t - 243)^{-1.01}\ ,
\end{equation}
and for the line energy:
\begin{equation}
    E_{\rm line} \propto (t - 243)^{-0.70}.
\end{equation}

\noindent We performed likelihood ratio tests to compare models with and without an intrinsic dispersion term, and found that including dispersion significantly improves the fit in all cases. The tests consistently rejected the null hypothesis of zero intrinsic dispersion, and thus the inclusion of dispersion as detailed above is therefore warranted.

\noindent The relationship between \( L_{\rm obs} \) and \( E_{\rm line} \) is fundamental in understanding their time evolution. For instance, when we first fit \( E_{\rm line} \) with the method described in \cite{2024SCPMA..6789511Z}, we obtain the power-law relations \( E_{\rm line} \propto (t - t_0)^{-1} \) and \( L_{\rm obs} \propto (t - t_0)^{-2} \), where \( t_0 = 223 \). Nevertheless, the time trends so derived are contrary to the given time-invariant relation \( L_{\rm obs} \propto E_{\rm line}^{1.47} \). Alternatively, by first finding \( L_{\rm obs} \) against time and subsequently finding \( E_{\rm line} \) against time, we are able to recover the \( L_{\rm obs} \)–\( E_{\rm line} \) relationship correctly. From a statistical viewpoint, this method provides a better fit than that presented in \cite{2024SCPMA..6789511Z}.

\subsection{Survival from Photodisintegration}

The detection of narrow MeV lines during the prompt emission of GRB 221009A has been interpreted as arising from Doppler-boosted gamma-ray transitions of $^{56}$Ni entrained in the relativistic jet. To determine whether these nuclei can survive the prompt radiation field without being destroyed by photodisintegration, we evaluate the photon number density and destruction timescale in the jet’s comoving frame.

\textcolor{black}{We take the interval from 290 to 300~s as a representative case for checking the photodisintegration process. Its spectrum is fitted with a Band function characterized by parameters \( A = 0.318 \), \( E_{\mathrm{p}} = 1014~\mathrm{keV} \), \( \alpha = -1.685 \), and \( \beta = -2.491 \). By integrating this function, we obtain a luminosity of approximately \( L_{\rm iso} \sim 10^{51}\,\mathrm{erg\,s^{-1}} \). A Lorentz factor $\Gamma \sim 50$ is inferred from the observed line energy, and a radius of \( R = 10^{14}~\mathrm{cm} \) is adopted to maintain consistency with the previous transparency calculation. The total comoving photon number density is then}
\begin{equation}
n’_\gamma \approx \frac{L_{\rm iso}}{4\pi R^2 \Gamma^2 c \langle \epsilon_{\rm obs} \rangle}
\approx \textcolor{black}{2.4 \times 10^{15}}~\mathrm{cm^{-3}}\ ,
\end{equation}
\textcolor{black}{where the mean observed photon energy, $\langle \epsilon_{\rm obs} \rangle = 27  ~\mathrm{keV}$, is derived from the Band spectrum.}

The photodisintegration threshold is typically $E^{\rm th} \sim 10~\mathrm{MeV}$ in the nucleus rest frame \cite{Wang:2007xj, Horiuchi:2012by}, this corresponds to the observer's frame
\textcolor{black}{
\begin{equation}
E_{\rm obs}^{\rm th}=\frac{\mathcal{D} }{1+z} E^{\rm th} 
             \simeq \textcolor{black}{0.86}\ {\rm GeV}.
\end{equation}
}
Integrating the Band function over photon energies above this comoving threshold gives a fractional abundance of high-energy photons:

\textcolor{black}{
\begin{equation}
f(E > E_{\rm obs}^{\rm th}) \approx \textcolor{black}{1.6\times10^{-7}}.
\end{equation}
}
Hence, the number density of photons above the threshold is:
\textcolor{black}{
\begin{equation}
n’_\gamma(E’ > E^{\rm th}) \approx f n’_\gamma \approx \textcolor{black}{3.9\times10^{8}}\ {\rm cm^{-3}}.
\end{equation}
}
The photodisintegration timescale can be estimated using
\textcolor{black}{
\begin{equation}
t_{A\gamma} \approx \frac{1}{n’_\gamma \sigma_{A\gamma} c} \approx \textcolor{black}{8.4 \times 10^{6}}~\mathrm{s},
\end{equation}
}
where we adopt a cross-section $\sigma_{A\gamma} \sim 10^{-26}~\mathrm{cm^2}$ \cite{1976ApJ...205..638P, 1993APh.....1..229K}. This is much longer than the dynamical timescale of the jet
\textcolor{black}{
\begin{equation}
\textcolor{black}{t_{\text{dyn}} = \frac{R}{\Gamma  c} = 66~\mathrm{s}.}
\end{equation}
}

Thus, the condition $t_{A\gamma} \gg t_{\text{dyn}}$ is clearly satisfied, confirming that photodisintegration is negligible during the relevant phase of jet propagation. {\color{black} It is worth to mention that we adopt a Lorentz factor of $\Gamma \approx 50$, consistent with modeling of GRB~221009A, as a representative mean value. Variations of $\Gamma$ within $\pm 50\%$ change the exact Doppler boost but do not affect the overall conclusions. Similarly, the assumed dissipation radius of $\sim 10^{14}$~cm can vary by factors of a few without substantially altering the agreement between the model predictions and the observed emission (Figure~\ref{fig:optical-depth} (b)).
}

\subsection{Survival from Nuclear Spallation}

Collisions with protons and neutrons in the flow can fragment nuclei through nuclear spallation.
Because both the baryons and the $^{56}$Ni clumps move with the same bulk Lorentz factor, the relevant frame is again the jet comoving frame. For an isotropic–equivalent kinetic power $L_{\rm iso}$ the comoving proton density is

\begin{equation}
n'_{p}\;=\;
\frac{L_{iso}}
     {4\pi R^{2}\,\Gamma^{2}\,m_{p}c^{3}} \simeq \textcolor{black}{7.1\times10^{10}}\ {\rm cm^{-3}},
\end{equation}
by taking the same parameters used for the photodisintegration analysis, that \textcolor{black}{ $L_{iso} \approx 10^{51}\,\mathrm{erg\,s^{-1}}$, 
$R = 10^{14}~ {\rm cm}$, and $\Gamma = 50$.} 

Laboratory data for proton collisions at a few hundred MeV per nucleon give an inelastic (fragmentation) cross-section $\sigma_{\rm sp}\;\simeq\;(3\text{–}5)\times10^{-25}\ {\rm cm^{2}}$ \cite{1983ASIC..107..321S, soppera2012janis}. The mean time between destructive collisions in the comoving frame is therefore

\begin{equation}
t_{\rm sp} = \frac{1}{\,n'_{p}\,\sigma_{\rm sp}\,c} \approx\; \textcolor{black}{1200}~ {\rm s}.
\end{equation}

Hence $t_{\rm sp} \gg t_{\rm dyn}$. The nuclear spallation is  too slow to deplete the entrained $^{56}$Ni before the flow expands and the density drops further. \textcolor{black}{
The selected value of $\Gamma$ corresponds to the mid/late prompt outflow component
analyzed here, which is physically distinct from the earliest, ultra-relativistic
prompt emission that dominates the extreme isotropic-equivalent energy of
GRB~221009A. As such, adopting $\Gamma \simeq 50$ does not contradict the BOAT
nature of the burst. We note that this choice is conservative: for larger Lorentz
factors, the proton density decreases as $n'_p \propto \Gamma^{-2}$,
while the dynamical time increases as $t_{\rm dyn} \sim R/(\Gamma c)$.
Consequently, higher values of $\Gamma$ further lengthen the ratio
$t_{\rm sp}/t_{\rm dyn}$, making nuclear spallation even less efficient.
The same considerations regarding the choice of Lorentz factor and its
consistency with the BOAT nature of GRB~221009A apply equally to the
photodisintegration analysis.
}

\subsection{Conditions for Neutrino Observation}

The electron-capture decay of \({}^{56}\text{Ni}\) produces neutrinos with a characteristic energy of \(Q_{\nu} = 0.41\,\text{MeV}\) in the nucleus's rest frame \citep{1994ApJS...92..527N}. However, when these neutrinos are emitted from a relativistic jet, their energies are Doppler-boosted. For a jet with a Lorentz factor of \(\Gamma \approx 150\), the effective observed neutrino energy can be estimated by applying a Doppler boost factor (approximately \(2\Gamma\) for an on-axis source), yielding 
\[
E_{\nu,\mathrm{obs}} \approx{\color{black}\frac{1}{1+z}} 0.41\,\text{MeV} \times 2\Gamma \approx  107\,\text{MeV}.
\]
Thus, the neutrinos are supposed to be observed at energies on the order of \(100\,\text{MeV}\). For GRB 221009A at \(z=0.151\) (luminosity distance \({\color{black}D}_L \approx 745\,\text{Mpc}\)), the neutrino luminosity is calculated to be 
\[
L_{\nu} = 2.5 \times 10^{49} \, \mathrm{erg\,s^{-1}},
\]
which corresponds to an observed flux of 
\[
F_{\nu,\mathrm{obs}} = \frac{L_{\nu}}{4 \pi {\color{black}D}_L^2} \approx 3.9 \times 10^{-7}\,\text{erg\,cm}^{-2}\,\text{s}^{-1}.
\]

Current MeV neutrino detectors, such as Super-Kamiokande and JUNO, have sensitivities of roughly \(10^{-5}\)–\(10^{-4}\,\text{erg\,cm}^{-2}\,\text{s}^{-1}\) for short-duration bursts. Even after accounting for the Doppler boost to approximately \(100\,\text{MeV}\) in the observer frame, the predicted neutrino flux of \(\sim 4 \times 10^{-7}\,\text{erg\,cm}^{-2}\,\text{s}^{-1}\) remains below the detection thresholds of these instruments. 

Similarly, high-energy neutrino observatories such as IceCube, which target neutrinos in the TeV--PeV range
\textcolor{black}{and are therefore insensitive to MeV--sub-GeV neutrinos}
\textcolor{black}{\citep{2017JInst..12P3012A,2017ApJ...843..112A}},
have not reported any significant signal from this event.
Future detectors, such as Hyper-Kamiokande and DUNE, are expected to improve MeV neutrino sensitivity
\textcolor{black}{to transient neutrino fluences} \textcolor{black}{\citep{2018PTEP.2018f3C01H,2020arXiv200203005A}}.
Under these improved conditions, similar GRB neutrino emission might become detectable.
Therefore, even though the neutrinos produced by $^{56}\text{Ni}$ decay (0.41~MeV in the rest frame)
are boosted to around \textcolor{black}{the $\sim$10--100~MeV range} in the observer frame,
the corresponding flux for GRB~221009A is still below the sensitivity thresholds of current MeV neutrino detectors.

\begin{redsubsection}{Energy-dependent Doppler Boosting of ${}^{56}{\rm Ni}$ Decay Lines: Constraints From Escape-Surface Effects} 

From the observation, the 158.38 keV and 269.50 keV lines in the jet comoving frame are detected at 12.22 MeV and 24.24 MeV at $\sim 300$~s in the observer frame. Using the Doppler relation $E_{\rm obs}=\mathcal{D} E'/(1+z)$ this implies $\mathcal{D}_{\vsim_1}=77.16(1+z)$ and $\mathcal{D}_{\vsim_2}=89.94(1+z)$, here we denote the two lines by $\vsim_1$ (158.38~keV) and $\vsim_2$ (269.50~keV). So the ratio $\mathcal{D}_{\vsim_2}/\mathcal{D}_{\vsim_1}=1.16$ is independent of redshift and shows that the 269.50 keV line is boosted by 16\% more than the 158.38 keV line. While a detailed treatment of nuclear physics in a relativistic jet is not yet settled, basic considerations suggest this mismatch is difficult to attribute to ionization or intrinsic nuclear physics, since these do not obviously shift the comoving rest energies, and instead points to the two lines becoming observable under different outflow conditions, for example different effective escape surface due to energy-dependent opacity.

Assume the lines are produced by ${}^{56}{\rm Ni}$ decay in the comoving frame, but the observed narrow ``line core'' is set at the last-scattering surface where $\tau(E)\sim 1$. This assumption applies if the line-forming region lies close to the scattering photosphere; 
if the region is already optically thin, the Doppler-factor mismatch must instead arise 
primarily from geometric or kinematic effects. The scattering opacity is dominated by Klein--Nishina Compton scattering \cite{1929ZPhy...52..853K}, so the effective cross section depends on photon energy in the electron rest frame. For $E'_{\vsim_1}$ and $E'_{\vsim_2}$,

\begin{equation}
\sigma_{\rm KN}(E'_{\vsim_1})\simeq 0.66\,\sigma_T,\qquad
\sigma_{\rm KN}(E'_{\vsim_2})\simeq 0.55\,\sigma_T,
\end{equation}
so
\begin{equation}
\frac{\sigma_{\rm KN}(E'_{\vsim_1})}{\sigma_{\rm KN}(E'_{\vsim_2})}\simeq 1.19.
\end{equation}

In a steady relativistic outflow with $n'(r)\propto r^{-2}$, the radial optical depth scales approximately as $\tau(E,r)\propto \sigma_{\rm KN}(E)/r$. Then the decoupling radius obeys $r_{\rm esc}(E)\propto \sigma_{\rm KN}(E)$, hence
\begin{equation}
\frac{r_{\rm esc, \vsim_1}}{r_{\rm esc, \vsim_2}}\simeq 1.19.
\end{equation}

So the 158~keV line decouples from the outflow, that is its line core is set by the last-scattering surface, at a radius about 19\% larger than for the 269.5~keV line. Using the standard Doppler-factor dependence
\begin{equation}
\mathcal{D}(\Gamma,\theta)=\frac{1}{\Gamma(1-\beta\cos\theta)}\simeq \frac{2\Gamma}{1+\Gamma^2\theta^2},
\end{equation}
the observed Doppler-factor difference between the two lines can be naturally interpreted as arising from differing conditions at their respective escape surfaces. In particular, the mismatch may reflect differences in $\Gamma$, $\theta$, or some combination of the two.

\textbf{A) Same $\theta$, different $\Gamma$ (different bulk speeds at different escape radii)} Because the Klein--Nishina cross section is smaller at higher photon energy, the photons associated with $\vsim_2$ decouple deeper in the flow (smaller $r_{\rm esc}$), where the jet is still faster, while the photons associated with $\vsim_1$ remain trapped longer and escape farther out, after the flow has relatively slowed. Quantitatively, assuming $\Gamma(r)\propto r^{-g}$, one has (for nearly fixed $\theta$) $\mathcal{D}\propto\Gamma$ and
\begin{equation}
\frac{\mathcal{D}_{\vsim_2}}{\mathcal{D}_{\vsim_1}}\approx \left(\frac{r_{\rm esc, \vsim_1}}{r_{\rm esc, \vsim_2}}\right)^{g}=1.19^{\,g},
\end{equation}
which gives $g=\ln(1.17)/\ln(1.19)\simeq0.89$, that is 
\begin{equation}
\Gamma(r)\propto r^{-0.89}
\end{equation}
across the $\sim 19\%$ separation in escape radius at $\sim 300$~s. For $z=0.151$, the inferred Doppler factors are $\mathcal{D}_{\vsim_1}=89$ and $\mathcal{D}_{\vsim_2}=100$, so if the line cores are close to on-axis (so that $\mathcal{D}\simeq 2\Gamma$) the implied Lorentz factors are $\Gamma_{\vsim_1}\simeq 44$ and $\Gamma_{\vsim_2}\simeq 52$.

\textbf{B) Same $\Gamma$, different effective $\theta$ (energy-dependent last-scattering angles).}
If the bulk Lorentz factor is essentially the same, the Doppler-factor mismatch can instead arise from differences in the effective escape-angle distribution at the last-interaction (decoupling) surface. In an energy-dependent radiative transfer problem, the effective opacity can vary with photon energy (e.g., through Klein--Nishina suppression and/or energy-dependent $\gamma\gamma$ opacity), so that the higher-energy photons associated with $\vsim_2$ can decouple deeper and with a more forward-peaked angular distribution, whereas the lower-energy photons associated with $\vsim_1$ decouple at slightly larger effective optical depth and therefore sample a modestly broader range of escape angles. This biases the effective $\theta$ for the photons associated with $\vsim_1$ to larger values and reduces $\mathcal{D}$ through the $(1+\Gamma^2\theta^2)^{-1}$ factor, without requiring multiple scatterings that would substantially broaden a narrow spectral feature. At fixed $\Gamma$, the required angle is obtained by inverting the small-angle form,

\begin{equation}
\theta(\Gamma,\mathcal{D})=\frac{1}{\Gamma}\sqrt{\frac{2\Gamma}{\mathcal{D}}-1}.
\end{equation}
Adopting the assumption that the photons associated with $\vsim_2$ escape on the line of sight, $\theta_{\vsim_2}=0$, its Doppler factor fixes the Lorentz factor at that escape surface via $\mathcal{D}_{\vsim_2}=\Gamma(1+\beta)\simeq 2\Gamma$. For $z=0.151$ we have $\mathcal{D}_{\vsim_2}=89.94(1+z)=103.53$, which implies $\Gamma=51.77$. Keeping the same $\Gamma$ for the $\vsim_1$ line, the observed $\mathcal{D}_{\vsim_1}=77.16(1+z)=88.81$ requires a nonzero effective escape angle $\theta_{\vsim_1}=0.45^\circ$. In this picture, the entire Doppler-factor mismatch is accounted for by a modest increase of the effective last-scattering angle of the photons associated with $\vsim_1$ relative to the on-axis photons associated with $\vsim_2$, consistent with stronger angular diffusion at lower photon energy.

The line-forming region may lie in the transition from the prompt emission to the onset of the afterglow, where the flow can still be energized. Here we compare our results with the Blandford–McKee solution \citep{1976PhFl...19.1130B}, but we note that the external shock self-similar conditions may not yet apply, so any use of Blandford–McKee scalings should be regarded as approximate. With this caveat, in case A the required gradient $\Gamma(r)\propto r^{-0.89}$ can be compared to the BM self-similar deceleration of an adiabatic relativistic blast wave in an external density profile $n(r)\propto r^{-k}$, for which $\Gamma(r)\propto r^{-(3-k)/2}$; identifying $g=(3-k)/2=0.89$ gives $k\simeq 1.2$, that is an external stratification between the standard ISM ($k=0$) and wind ($k=2$) limits. Therefore the $\Gamma(r)$ scaling suggested by the line ratio is not excluded by BM-type dynamics, but it would point either to an intermediate density profile or to an effective stratification produced by jet structure and/or additional dissipation. In case B, the inferred $\theta_{\vsim_1}=0.45^\circ$ is smaller than typical GRB jet opening angles, so an escape-angle bias of this magnitude is geometrically allowed and can arise from a modest broadening of the escape angle distribution. In practice, the observed Doppler-factor mismatch need not be purely A or purely B: a combined scenario in which $\vsim_2$ decouples at smaller radius with slightly larger $\Gamma$ and also with a more forward-peaked angle distribution is physically natural in this energy-dependent radiative transfer problem.

Overall, the different last-scattering surfaces provide a self-consistent explanation of the observed 16\% energy mismatch between the two lines.

\end{redsubsection}

{\bf Data availability}

The GECAM data supporting this work can be accessed via the National Space Science Data Center (NSSDC) at \url{https://gecam.nssdc.ac.cn/}. The Fermi satellite data used in this study are publicly available from the Fermi Science Support Center (FSSC) at \url{https://fermi.gsfc.nasa.gov/ssc/data/}.  

{\bf Code Availability}

The GECAM data reduction was performed using the standard software GECAMTools available at the GitHub website \url{https://github.com/zhangpeng-sci/GECAMTools-Public}. The analysis in this work was performed using the standard Fermi Science Tools, which are publicly available at \url{https://fermi.gsfc.nasa.gov/ssc/data/analysis/software/}. 
\clearpage

\begin{addendum}

 \item[Acknowledgements] 
 R. Moradi acknowledges support from the Institute
of High Energy Physics, Chinese Academy of Sciences
(E32984U810). E.S.Yorgancioglu acknowledges support from the “Alliance of International
Science Organization (ANSO) Scholarship For Young Talents.” We acknowledge support from 
the Strategic Priority Research Program of the Chinese Academy of Sciences (Grant No. XDB0550300% xiongshaolin(taolian), HXMT+GECAM
), the National Key R\&D Program of China (2021YFA0718500), %xiongshaolin
the National Natural Science Foundation of China (grant Nos. 12273042,%xiong shaolin, GECAM
12494572, 12494570,%xiong shaolin, SVOM
12333007%zhang shuangnan
), and China's Space Origins Exploration Program. %zhang shuangnan et al., eXTP
The authors gratefully acknowledge insightful discussions regarding the model parameters with S. Nagataki and J. A. Rueda.
 
\item[Author Contributions]  
R.M. conceived and initiated the project, developed the theoretical framework and models, performed key calculations, and led the manuscript preparation. E.S.Y. carried out the joint luminosity–time fitting, refined the model parameters, contributed to the interpretation of results, and co-wrote the manuscript. S.-L.X. supplied and curated the GECAM dataset, contributed to its interpretation, and helped structure the manuscript. Y.-Q.Z. conducted the detailed GECAM data analysis and assisted with advanced spectral diagnostics. S.-N.Z. participated in scientific discussions, contributed to manuscript writing, and performed cross-checks on all interpretations. R.D. proposed the extended search for additional spectral lines and guided the spectral analysis strategy. Y.W. co-initiated the project, discovered the second MeV line, and participated in the theoretical modeling and writing. All authors discussed the results and edited the manuscript.

\item[Competing Interests] The authors declare no competing interests.

\item[Correspondence] Correspondence and requests for materials should be addressed to  rmoradi@ihep.ac.cn,
emre@ihep.ac.cn,
xiongsl@ihep.ac.cn,
yu.wang@icranet.org.

\end{addendum}

\begin{table}
\centering
\caption{Gamma Transitions for \(^{56}\text{Ni}\) Decay \citep{sur1990reinvestigation, 1994ApJS...92..527N}.} 
\label{tab:ni_decay}
\begin{tabular}{cc}
\hline
\hline
$\rm Energy~(MeV)$ & $\rm Intensity~ (\%)$ \\
\hline
$0.158$ & $98.8$ \\
$0.270$ & $36.5$ \\
$0.480$ & $36.6$ \\
$0.750$ & $49.5$ \\
$0.812$ & $86.0$ \\
$1.560$ & $14.0$ \\
\hline
\end{tabular}
\end{table}

\begin{table}
\centering
\caption{The spectral line component, including the observed time interval, flux, and corresponding isotropic luminosity for GRB 221009A, obtained from GECAM-C and \emph{Fermi}/GBM. The data were {taken} from Zhang et al \cite{2024SCPMA..6789511Z} with the permission of the authors. }
\label{tab:MeV_Lines}
\begin{tabular}{c|c|c}
\hline
\hline
$\rm Time~{\color{black} interval}~(s)$ &  $\rm Flux~(\mathrm{erg\,cm^{-2}\,s^{-1}}$) &  $\rm Luminosity\,(\mathrm{erg\,s^{-1}})$ \\
\hline
\hline
$(246, 256)$ & $({2.02}_{-0.06}^{+2.67}) \times 10^{-5}$ & $({1.27}_{-0.04}^{+1.67}) \times 10^{51}$ \\ \hline
$(270, 275) $ & $ ({4.02}_{-2.04}^{+1.76}) \times 10^{-6}$ & $({2.52}_{-1.27}^{+1.10}) \times 10^{50}$ \\ \hline
$(275, 280) $ & $ ({3.32}_{-0.72}^{+1.23}) \times 10^{-6}$ & $({2.08}_{-0.45}^{+0.77}) \times 10^{50}$ \\ \hline
$(280, 285) $ & $ ({3.98}_{-0.28}^{+3.42}) \times 10^{-6}$ & $({2.50}_{-0.18}^{+2.14}) \times 10^{50}$ \\ \hline
$(285, 290) $ & $ ({3.31}_{-0.60}^{+1.55}) \times 10^{-6}$ & $({2.07}_{-0.37}^{+0.97}) \times 10^{50}$ \\ \hline
$(290, 295) $ & $ ({2.94}_{-0.51}^{+1.37}) \times 10^{-6}$ & $({1.84}_{-0.32}^{+0.86}) \times 10^{50}$ \\ \hline
$(295, 300) $ & $ ({1.56}_{-0.24}^{+0.72}) \times 10^{-6}$ & $({0.98}_{-0.15}^{+0.45}) \times 10^{50}$ \\ \hline
$(280, 300) $ & $ ({2.62}_{-0.33}^{+0.53}) \times 10^{-6}$ & $({1.64}_{-0.21}^{+0.33}) \times 10^{50}$ \\ \hline
$(300, 310) $ & $ ({1.74}_{-0.47}^{+0.51}) \times 10^{-6}$ & $({1.09}_{-0.29}^{+0.32}) \times 10^{50}$ \\ \hline
$(310, 320) $ & $ ({1.58}_{-0.31}^{+1.02}) \times 10^{-6}$ & $({0.99}_{-0.19}^{+0.64}) \times 10^{50}$ \\ \hline
$(300, 320) $ & $ ({1.51}_{-0.25}^{+0.39}) \times 10^{-6}$ & $({0.95}_{-0.16}^{+0.24}) \times 10^{50}$ \\ \hline
$(320, 340) $ & $ ({1.40}_{-0.59}^{+0.04}) \times 10^{-6}$ & $({0.88}_{-0.37}^{+0.03}) \times 10^{50}$ \\ \hline
$(340, 360) $ & $ ({1.60}_{-0.12}^{+0.54}) \times 10^{-6}$ & $({1.00}_{-0.08}^{+0.34}) \times 10^{50}$ \\ \hline
$(320, 360) $ & $ ({1.39}_{-0.24}^{+0.52}) \times 10^{-6}$ & $({0.87}_{-0.15}^{+0.33}) \times 10^{50}$ \\ \hline
\hline
\end{tabular}
\end{table}

\begin{table}
\centering
\caption{The spectral line component, including the observed time interval, and corresponding isotropic luminosity for GRB 221009A, obtained from \emph{Fermi}/GBM. {\color{black} The luminosity values are the median of the marginalized posterior distributions, and the errors denote symmetric 1-$\sigma$ uncertainties.} The data were {taken} from Ravasio et al. (2024) \citep{2024Sci...385..452R}.}
\begin{tabular}{c|c}
\hline
$\rm Time  ~ {\color{black} interval}~ (s)$ & $\rm Luminosity~ (\,\mathrm{erg\,s^{-1}})$ \\
\hline
${\color{black}(280,300)}$ & $(1.12^{+ 0.19}_{- 0.19}) \times 10^{50}$ \\
${\color{black}(280,285)}$  & $(0.77^{+ 0.42}_{- 0.42}) \times 10^{50}$ \\
${\color{black}(285,290)}$  & $(0.43^{+ 0.33}_{- 0.28}) \times 10^{50}$ \\
${\color{black}(290,295)}$  & $(1.84^{+ 0.36}_{- 0.33}) \times 10^{50}$ \\
${\color{black}(295,300)}$  & $(0.63^{+ 0.28}_{- 0.27}) \times 10^{50}$ \\
${\color{black}(300,320)}$ & $(1.14^{+ 0.20}_{- 0.18}) \times 10^{50}$ \\
${\color{black}(300,310)}$ & $(1.08^{+ 0.19}_{- 0.17}) \times 10^{50}$ \\
${\color{black}(310,320)}$  & $(0.75^{+ 0.21}_{- 0.19}) \times 10^{50}$ \\
${\color{black}(320,340)}$ & $(0.23^{+ 0.15}_{- 0.13}) \times 10^{50}$ \\
${\color{black}(340,360)}$ & $(0.21^{+ 0.12}_{- 0.10}) \times 10^{50}$ \\
\hline
\end{tabular}
\label{tab:Lgauss}
\end{table}

\begin{table*}[ht!]
\centering
\caption{\textcolor{black}{Model comparison for the $290-300$~s spectrum. We list the Deviance Information Criterion (DIC) and Widely Applicable Information Criterion (WAIC) for the one-line (Band + G) and two-line (Band + G + G) models, along with differences and the approximate Bayes factor derived from $\Delta$ WAIC. Lower DIC/WAIC indicates a better balance of fit quality and model complexity.}}
\label{tab:model_comparison}
\begin{tabular}{lccccc}
\hline
Model & DIC & WAIC & $\Delta$DIC & $\Delta$WAIC & Bayes Factor \\
\hline
Band+G+G & 141.14 & 169.58 & --- & --- & --- \\
Band+G & 145.48 & 174.14 & 4.34 & 4.56 & 0.10 \\
\hline
\end{tabular}
\end{table*}

\end{document}